\documentclass[preprints,article,accept,moreauthors,pdftex]{Definitions/mdpi}
\firstpage{1} 
\pubvolume{xx}
\issuenum{1}
\articlenumber{5}
\pubyear{2020}
\copyrightyear{2020}
\history{ }
\graphicspath{{./figures/}}
\usepackage[utf8]{inputenc}
\usepackage[english]{babel}

\usepackage{amsfonts}
\usepackage{mathrsfs}
\usepackage{mathtools} %extension of amsmath, provides dcases

\usepackage{tabularx}
\usepackage{physics,subfigure}
\usetikzlibrary{graphs,graphs.standard,fit,shapes.geometric,calc,hobby}

\allowdisplaybreaks
\Title{Transport efficiency of continuous-time quantum walks on graphs}

\Author{Luca Razzoli$^{1,\ast}$, Matteo G. A. Paris$^{2,3}$, and Paolo Bordone$^{1,4,\ast}$}

\AuthorNames{Luca Razzoli, Matteo G. A. Paris and Paolo Bordone}

\address{%
$^{1}$ \quad Dipartimento di Scienze Fisiche, Informatiche e 
Matematiche, Universit\`{a} di Modena e Reggio Emilia, I-41125 Modena, 
Italy\\
$^{2}$ \quad Quantum Technology Lab, Dipartimento di Fisica {\em Aldo Pontremoli}, Universit\`{a} degli Studi di Milano, I-20133 Milano, Italy\\
$^{3}$ \quad INFN, Sezione di Milano, I-20133 Milano, Italy\\
$^{4}$ \quad Centro S3, CNR-Istituto di Nanoscienze, I-41125 Modena, Italy
}

\corres{Correspondence: luca.razzoli@unimore.it (L.R.); paolo.bordone@unimore.it (P.B.)}
\abstract{Continuous-time quantum walk describes the propagation of a quantum particle (or an excitation) evolving continuously in time on a graph. As such, it provides a natural framework for modeling transport processes, e.g., in light-harvesting systems. In particular, the transport properties strongly depend on the initial state and on the specific features of the graph under investigation.  In this paper, we address the role of graph topology, and investigate the transport properties of graphs with different regularity, symmetry, and connectivity. We neglect disorder and decoherence, and assume a single trap vertex accountable for the loss processes. In particular, for each graph, we analytically determine the subspace of states having maximum transport efficiency. Our results provide a set of benchmarks for environment-assisted quantum transport, and suggest that connectivity is a poor indicator for transport efficiency. Indeed, we observe some specific correlations between transport efficiency and connectivity for certain graphs, but in general they are uncorrelated.}

  % List three to ten pertinent keywords specific to the article, yet reasonably common within the subject discipline.

\begin{document}
%%%%%%%%%%%%%%%%%%%%%%%%%%%%%%%%%%%%%%%%%%

%The order of the section titles is: Introduction, Materials and Methods, Results, Discussion, Conclusions for these journals: aerospace,algorithms,antibodies,antioxidants,atmosphere,axioms,biomedicines,carbon,crystals,designs,diagnostics,environments,fermentation,fluids,forests,fractalfract,informatics,information,inventions,jfmk,jrfm,lubricants,neonatalscreening,neuroglia,particles,pharmaceutics,polymers,processes,technologies,viruses,vision

%%%%%%%%%%%%%%%%%%%%%%%%%%%%%%%%%%%%%%%%%%

\section{Introduction}
%CTQW
A continuous-time quantum walk (CTQW) is the quantum mechanical counterpart of the continuous-time random walk. It describes the dynamics of a quantum particle which evolves continuously in time in a discrete space, e.g. on the vertices of a graph, obeying the Schr\"odinger equation \cite{farhi1998quantum,childs2002example}. The Hamiltonian describing a CTQW is usually the Laplacian matrix $L$, which encodes the topology of the graph and plays the role of the kinetic energy of the walker. Experimentally \cite{wang2013physical}, CTQWs can be implemented on nuclear-magnetic-resonance quantum computers \cite{du2003experimental}, optical lattices of ultracold Rydberg atoms \cite{cote2006quantum}, quantum processors \cite{qiang2016efficient}, and photonic chips \cite{tang2018experimental}.
%APPLICATIONS
Applications of CTQWs range from implementing fast and efficient quantum algorithms \cite{venegas2008quantum,portugal2018quantum}, e.g. for spatial search \cite{childs2004spatial} and image segmentation \cite{krok2019application}, to implementing quantum logic gates by multi-particle CTQWs in 1D \cite{lahini2018quantum}, from universal computation \cite{childs2009universal} to modeling and simulating quantum phenomena, e.g. state transfer \cite{christandl2004perfect,kendon2011perfect,alvir2016perfect}, quantum transport, and for characterizing the behavior of many-body systems \cite{lahini2012quantum,beggi2018probing}.

%QUANTUM TRANSPORT
Modeling quantum transport processes by means of CTQWs is indeed a well-established practice and an appropriate mathematical framework. 
Quantum transport has been investigated with this approach on restricted geometries \cite{agliari2008dynamics}, semi-regular spidernet graphs \cite{salimi2010continuous}, Sierpinski fractals \cite{darazs2014transport}, and on large-scale sparse regular networks \cite{li2020quantum}. CTQWs have been used to model transport of nonclassical light in coupled waveguides \cite{rai2008transport}, coherent exciton transport on hierarchical systems \cite{blumen2006coherent}, small-world networks \cite{mulken2007quantum}, Apollonian networks \cite{xu2008coherent}, and on an extended star graph \cite{yalouz2018continuous}, coherent transport on complex networks \cite{mulken2011continuous}, and exciton transfer with trapping \cite{mulken2007survival,agliari2010continuous}. It is worth noting that CTQWs do not necessarily perform better than their classical counterparts, since the transport properties strongly depend on the graph, the initial state, and on the propagation direction under investigation \cite{mulken2005slow}. A measure of the efficiency of quantum and classical transport on graphs by means of density of states has been proposed in \cite{mulken2006efficiency}.

%QT in BIO
Biological systems are known to show quantum effects \cite{lambert2013quantum,mohseni2014quantum} and efficient transport processes. Hence the great interest in studying also CTWQs to model, e.g., exciton transport on dendrimers \cite{mulken2006coherent}, photosynthetic energy transfer \cite{mohseni2008environment}, environment-assisted quantum transport \cite{rebentrost2009environment}, dephasing-assisted transport on quantum networks and biomolecules \cite{plenio2008dephasing}, excitation transfer in light-harvesting systems \cite{olaya2008efficiency,caruso2009highly} and its limits \cite{hoyer2010limits}. There also studies concerning disorder-assisted quantum transport on hypercubes and binary trees \cite{novo2016disorder}, because the latter can model dendrimer-like structure for artificial light-harvesting systems \cite{adronov2000light,bradshaw2011mechanisms}.

%THIS WORK
A full characterization of the transport properties on different structures is therefore desired. Formally speaking, the CTQW Hamiltonian modeling transport processes shows similarities with the CTQW Hamiltonian adopted to study the spatial search. Both of them consist of the sum, with proper coefficients, of the Laplacian matrix, accountable for the motion of the walker on the graph, and the projector onto one or more specific vertices. This projector is the trapping Hamiltonian in transport problems and the oracle Hamiltonian in spatial search problems. Regularity, global symmetry, and connectivity of the graph have proved to be unnecessary for fast spatial search \cite{wong2016laplacian,janmark2014global,meyer2015connectivity} by invoking certain graphs, e.g. complete bipartite graphs, strongly regular graphs, joined complete graphs, and a simplex of complete graphs, as counterexamples of these false beliefs. In this work, we address the transport by CTQW on the above mentioned graphs, which are different in terms of regularity, symmetry, and connectivity, and we assess the transport efficiency for initial states localized at a vertex and for an initial superposition of two vertices. Our focus is on the role of connectivity, if any. Indeed, regularity and global symmetry are not required for efficient transport, because removing some edges in the complete graph and the hypercube, which are regular and highly symmetric graphs, has been shown to improve the transport efficiency \cite{novo2015systematic}. 

%STRUCTURE
The paper is organized as follows. In Sec. \ref{sec:ctqw} we introduce CTQWs on graph. In Sec. \ref{sec:dimrecmeth} we review the dimensionality reduction method to analyze CTQW problems \cite{novo2015systematic}, according to which we obtain a reduced model of the Hamiltonian encoding the problem considered and the reduced Hamiltonian still fully describes the dynamics relevant to the problem. In Sec. \ref{sec:qt} we define the Hamiltonian modeling the transport on graphs and the transport efficiency as figure of merit to measure the transport properties of the system. For each graph considered, we provide the reduced Hamiltonian and we compute the transport efficiency for different initial states. In Sec. \ref{sec:connectivity} we assess different measures of connectivity to characterize each graph considered. Finally, we present our conclusions in Sec. \ref{sec:conclusion}. In Appendix  \ref{app:subspace_equality} we report and refine the proof of the equality of the two subspaces required to compute the transport efficiency. In Appendix \ref{app:basis} we determine the basis states spanning such subspace for each graph considered.

%%%%%%%%%%%%%%%%%%%%%%%%%%%%%%%%%%%%%%%%%%
\section{Continuous-time quantum walks}
\label{sec:ctqw}
A graph is a pair $G=(V,E)$, where $V$ denotes the non-empty set of vertices and $E$ the set of edges. The order of the graph is the number of vertices, $\vert V \vert =N$. We define the adjacency matrix %($A_{jk}=1$ if the vertices $j$ and $k$ are connected, $0$ otherwise)

\begin{equation}
A_{jk}=\begin{cases}
1 & \text{if $(j,k)\in E$,}\\
0 & \text{otherwise,}
\end{cases}
\label{eq:adj_matr}
\end{equation}
which describes the connectivity of $G$, and $D$ the diagonal degree matrix with $D_{jj}=\operatorname{deg}(j)$, the degree of vertex $j$. In terms of these matrices, we introduce the graph Laplacian $L=D-A$, which is the matrix representation of the graph. According to this definition, $L$ is positive semidefinite, and singular.

The CTQW is the propagation of a quantum particle with kinetic energy when confined to a discrete space, e.g. a graph. The CTQW on a graph $G$ takes place on a $N$-dimensional Hilbert space $\mathcal{H}=\operatorname{span}(\{\ket{v} \mid v \in V\})$, and the kinetic energy term ($\hbar=1$) ${T}=-\nabla^2/2m$ is replaced by ${T}=\gamma L$, where $\gamma \in \mathbb{R}^+$ is the hopping amplitude of the walk. The state of the walker obeys the Schr\"{o}dinger equation

\begin{equation}
i\frac{d}{dt} \ket{\psi(t)}=H\ket{\psi(t)}\,,
\label{eq:schrodinger}
\end{equation}
with Hamiltonian $H=\gamma L$. Hence, a walker starting in the state $\ket{\psi_0}\in \mathcal{H}$ evolves continuously in time according to

\begin{equation}
\ket{\psi(t)}=U(t)\ket{\psi_0}\,,
\label{eq:psit}
\end{equation}
with $U(t)=\exp[-i H t]$ the unitary time-evolution operator. The probability to find the walker in a target vertex $w$ is therefore $\abs{\bra{w}\exp\left[-i H t\right]\ket{\psi_0}}^2$.

\section{Dimensionality reduction method}
\label{sec:dimrecmeth}
In most CTQW problems, the quantity of interest is the probability amplitude at a certain vertex of the graph. The graph encoding the problem to solve often contains symmetries which allow us to simplify the problem, since the evolution of the system actually occurs in a subspace of the complete $N$-dimensional Hilbert space $\mathcal{H}$ spanned by the vertices of the graph. We can determine the minimal subspace which contains the vertex of interest and is invariant under the unitary time evolution via the dimensionality reduction method for CTQW, proposed by Novo \textit{et al.} \cite{novo2015systematic}, which we briefly review in this section for completeness. Such subspace, also known as a Krylov subspace \cite{jafarizadeh2007investigation}, contains the vertex of interest and all powers of the Hamiltonian applied to it. The relevance and the power of this method is that the graph encoding a given problem can be mapped onto an equivalent weighted graph, whose order is lower than the order of the original graph and whose vertices are the basis states of the invariant subspace. The corresponding reduced Hamiltonian still fully describes the dynamics relevant to the considered problem. 

The unitary evolution  \eqref{eq:psit} can be expressed as

\begin{equation}
\ket{\psi(t)}=\sum_{k=0}^\infty \frac{(-it)^k}{k!}H^k\ket{\psi_0}\,,
\end{equation}
so $\ket{\psi(t)}$ is contained in the subspace $\mathcal{I}(H,\ket{\psi_0}) = \operatorname{span}(\lbrace H^k \ket{\psi_0} \mid k \in \mathbb{N}_0\rbrace)$. This subspace of $\mathcal{H}$ is invariant under the action of the Hamiltonian and thus also of the unitary evolution. Naturally, $\dim \mathcal{I}(H,\ket{\psi_0})\leq \dim \mathcal{H}=N$, but if the Hamiltonian is highly symmetrical, only a small number of powers of $H^k\ket{\psi_0}$ are linearly independent, so the dimension of $\mathcal{I}(H,\ket{\psi_0})$ can be much smaller than $N$.

Let $P$ be the projector onto $\mathcal{I}(H,\ket{\psi_0})$. Then

\begin{equation}
U(t) \ket{\psi_0}=P U(t) P\ket{\psi_0}=\sum_{k=0}^\infty \frac{(-it)^k}{k!}(PHP)^k\ket{\psi_0}=e^{-iPHPt}\ket{\psi_0}=e^{-iH_\textup{red}t}\ket{\psi_0}\,,
\end{equation}
where $H_\textup{red}=PHP$ is the reduced Hamiltonian, and we used the fact that $P^2=P$ (projector), $P\ket{\psi_0}=\ket{\psi_0}$, and $P U(t) \ket{\psi_0}= U(t) \ket{\psi_0}$.

For any state $\ket{\phi}\in\mathcal{H}$, which we consider the solution of the CTQW problem, we have

\begin{align}
\bra{\phi}U(t) \ket{\psi_0}&=\bra{\phi} P P U(t) P\ket{\psi_0}=\bra{\phi}P e^{-iH_\textup{red}t}\ket{\psi_0}
=\bra{\phi_\textup{red}} e^{-iH_\textup{red}t}\ket{\psi_0}\,,
\end{align}
where, the reduced state, $\ket{\phi_\textup{red}}=P\ket{\phi}$. Reasoning analogously with the projector $P'$ onto the subspace $\mathcal{I}(H,\ket{\phi})$ we obtain

\begin{equation}
\bra{\phi}U(t) \ket{\psi_0}=\bra{\phi} e^{-iH_\textup{red}'t}\ket{{\psi_0}_\textup{red}}\,,
\end{equation}
with $H_\textup{red}'=P'HP'$ and $\ket{{\psi_0}_\textup{red}}=P'\ket{\psi_0}$.

An orthonormal basis of $\mathcal{I}(H,\ket{\phi})$, denoted by $\{\ket{e_1},\ldots,\ket{e_m}\}$, can be obtained iteratively as follows: the first basis state is $\ket{e_1}=\ket{\phi}$, then the successive ones are obtained by applying $H$ on the current basis state and orthonormalizing with respect to the previous basis states. The procedure stops when we find the minimum $m$ such that $H\ket{e_m}\in \operatorname{span}(\{\ket{e_1},\ldots,\ket{e_m}\})$. The reduced Hamiltonian, i.e. $H$ written in the basis of the invariant subspace, has a tridiagonal form, so the original problem is mapped onto an equivalent problem governed by a tight-binding Hamiltonian of a line with $m$ sites.

%%%%%%%%%%%%%%%%%%%%%%%%%%%%%%%%%%%%%%%%%%
\section{Quantum transport}
\label{sec:qt}
The CTQW on a graph $G(V,E)$ of $N$ vertices provides a useful framework to model, e.g., the dynamics of a particle or a quasi-particle (excitation) in a network. The quantum walker moves under the Hamiltonian

\begin{equation}
H=\gamma L=\gamma \sum_{i\in V} \deg(i) \dyad{i}-\gamma\sum_{(i,j)\in E}(\vert i \rangle\langle j \vert + \vert j \rangle \langle i \vert)\,,
\label{eq:tbH}
\end{equation}
which can be read as a tight-binding Hamiltonian with uniform nearest-neighbor couplings $\gamma$ and on-site energies $\gamma \deg(i)$. In the following we set the units such that $\gamma = \hbar= 1$, so hereafter time and energy will be dimensionless.

However, in general, an excitation does not stay forever in the system in which it was created. In biological light-harvesting systems, the excitation gets absorbed at the reaction center, where it is transformed into chemical energy. In such scenario, the total probability of finding the excitation within the network is not conserved. We assume a graph in which the walker can only vanish at one vertex $w \in V$, known as \textit{trap vertex} or \textit{trap}. The component of the walker's wave function at the trap vertex is absorbed by the latter at a trapping rate $\kappa\in\mathbb{R}^+$ \cite{mulken2011continuous}. Then, to phenomenologically model such loss processes we have to change the Hamiltonian \eqref{eq:tbH}, so we introduce the trapping Hamiltonian

\begin{equation}
H_\textup{trap} = -i \kappa \dyad{w}\,,
\label{eq:trapH}
\end{equation}
which is anti-hermitian. This leads to the desired non-unitary dynamics described by the total Hamiltonian

\begin{equation}
H=L-i\kappa \dyad{w}\,.
\label{eq:qtH}
\end{equation}
This Hamiltonian has the same structure as the Hamiltonian for the spatial search of a marked vertex $w$ \cite{childs2004spatial}, i.e. it is the sum of the Laplacian matrix and the projector onto $\ket{w}$, with proper coefficients. For spatial search, the projector onto $\ket{w}$ plays the role of the oracle Hamiltonian and the search Hamiltonian is hermitian. For quantum transport, the projector onto $\ket{w}$, because of the pure imaginary constant, plays the role of the trapping Hamiltonian \eqref{eq:trapH} and the transport Hamiltonian \eqref{eq:qtH} is not hermitian.

A relevant measure for a quantum transport process is the transport efficiency \cite{rebentrost2009environment}, which can be defined as the integrated probability of trapping at the vertex $w$ 

\begin{equation}
\eta = 2 \kappa \int_0^{+\infty} \bra{w} \rho(t) \ket{w}\,dt=1-\Tr\left[ \lim_{t\to+\infty}\rho(t) \right]\,,
\label{eq:transport_eff_def}
\end{equation}
where $2\kappa \langle w \vert \rho(t) \vert w \rangle dt$ is the probability that the walker is successfully absorbed at the trap within the time interval $[t,t+dt]$ and $\rho(t)=\vert \psi(t) \rangle \langle \psi (t) \vert$ is the density matrix of the walker. The second equality of Eq. \eqref{eq:transport_eff_def} is due to the following reason. The surviving total probability of finding the walker within the graph at time $t$ is $\braket{\psi(t)}=\Tr[\rho(t)]$ and it is $\leq 1$ because of the loss processes at the trap vertex. Since the transport efficiency is the integrated probability of trapping in the limit of infinite time, we can also assess the transport efficiency as the complement to $1$ of the probability of surviving within the graph, which is the complementary event.

%we can also assess the transport efficiency as the probability of the complementary event of surviving within the graph.

\begin{figure}[tb!]
\centering
\begin{tikzpicture}[scale=0.4,use Hobby shortcut]
\begin{scope}[shift={(-10.5,0)}]
	\path
	(-7,0) coordinate (z0)
 	(-5,2) coordinate (z1)
	(-3,3) coordinate (z2)
	(0,2) coordinate (z3)
	(1,2) coordinate (z4)
	(5,3) coordinate (z5)
	(7,1) coordinate (z6)
	(8,-2) coordinate (z7)
	(0,-2) coordinate (z8)
	(-6,-3) coordinate (z9);
	\draw[line width=1.5pt,closed] (z0) .. (z1) .. (z2) .. (z3) .. (z4) .. (z5) .. (z6).. (z7).. (z8).. (z9);
	\draw[line width=1.5pt] (z4) .. (z8);
	
	\node[label=above:{\large $\mathcal{H}$}] at (1,2) {};
	\node[label=\text{$\braket{\psi_0}{\psi_0}=1$}] at (1,-5.5) {};        	
	
	%SUBSPACE-LEFT:
	\node[label=\textit{Subspace $\Lambda$}] at (-3,0) {};
	\node[label=$\ket{\psi_\Lambda}$] at (-3,-1.5) {};
	
	%TRAP: spiral
%	\draw [red] (-5,-2) circle [radius=0.025];
	\begin{scope}[shift={(-5,-2)}]
    \draw [domain=0:50,variable=\t,smooth,samples=500,draw=red]
        plot ({\t r}: {0.025+0.5*exp(-0.05*\t)});
    \end{scope}
	\node[label=right:\textcolor{red}{$\ket{w}$}] at (-5,-2) {};	

	%SUBSPACE-RIGHT:
	\node[label=\textit{Subspace $\bar{\Lambda}$}] at (4,0) {};
	\node[label=$\ket{\psi_{\bar{\Lambda}}}$] at (4,-1.5) {};

\end{scope}

\draw[-latex,line width=1.5pt] (-1.5,-1) -- (0,-1) node[above]{$t\to+\infty$} -- (1.5,-1);
\begin{scope}[shift={(9.5,0)}]
	\path
	(-7,0) coordinate (z0)
 	(-5,2) coordinate (z1)
	(-3,3) coordinate (z2)
	(0,2) coordinate (z3)
	(1,2) coordinate (z4)
	(5,3) coordinate (z5)
	(7,1) coordinate (z6)
	(8,-2) coordinate (z7)
	(0,-2) coordinate (z8)
	(-6,-3) coordinate (z9);
	\draw[line width=1.5pt,closed] (z0) .. (z1) .. (z2) .. (z3) .. (z4) .. (z5) .. (z6).. (z7).. (z8).. (z9);
	\draw[line width=1.5pt] (z4) .. (z8);

	\node[label=above:{\large $\mathcal{H}$}] at (1,2) {};
	\node[label=\text{$\braket{\psi(\infty)}{\psi(\infty)}\leq 1$}] at (1,-5.5) {};

	%SUBSPACE-LEFT
	\node[label=\textit{Subspace $\Lambda$}] at (-3,0) {};
%	\node[label=\text{$\ket{\psi_\Lambda(\infty)}=0$}] at (-3,-2) {};
	\node[label=\text{$0$}] at (-3,-1.5) {};
	
	%TRAP: spiral
%	\draw [red] (-5,-2) circle [radius=0.025];
	\begin{scope}[shift={(-5,-2)}]
    \draw [domain=0:50,variable=\t,smooth,samples=500,draw=red]
        plot ({\t r}: {0.025+0.5*exp(-0.05*\t)});
    \end{scope}
	\node[label=right:\textcolor{red}{$\ket{w}$}] at (-5,-2) {};

	%SUBSPACE-RIGHT
	\node[label=\textit{Subspace $\bar{\Lambda}$}] at (4,0) {};
	\node[label=$\ket{\psi_{\bar{\Lambda}}(\infty)}$] at (4,-1.5) {};

\end{scope}
\end{tikzpicture}
\caption{The quantum walker is in the initial state $\ket{\psi_0}$ \eqref{eq:psi0_Lambda_cmpnts} and has components in $\Lambda (H,\ket{w})$, the subspace spanned by the eigenstates of the Hamiltonian having a non-zero overlap with the trap $\ket{w}$, and in $\bar{\Lambda}=\mathcal{H}\setminus \Lambda$, the complement of $\Lambda$ in the complete Hilbert space $\mathcal{H}$. In the limit of $t \to +\infty$, the dynamics is such that the component having non-zero overlap with the trap is fully absorbed by the trap, i.e. $\ket{\psi_{\bar{\Lambda}}(\infty)}=0$, whereas the component in $\bar{\Lambda}$ survives. The dynamics is not unitary and the total survival probability of finding the walker within the graph is not conserved, i.e. $\braket{\psi(\infty)}{\psi(\infty)}\leq 1$.}
\label{fig:nonunitary_dyn}
\end{figure}
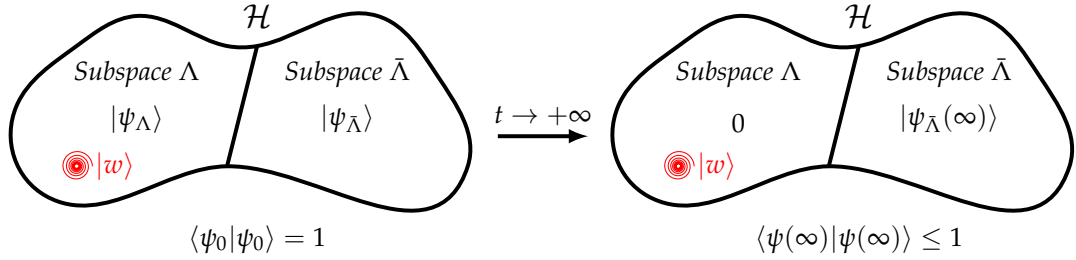

In this scenario there is no disorder in the couplings or site energies of the Hamiltonian nor decoherence during the transport.  In this ideal regime computing the transport efficiency amounts to finding the overlap of the initial state with the subspace $\Lambda(H,\ket{w})$ spanned by the eigenstates of the Hamiltonian $\ket{\lambda_k}$ having a non-zero overlap with the trap $\ket{w}$, as proved by Caruso \textit{et al.} \cite{caruso2009highly}. Indeed, the dynamics is such that the component of the initial state within the space $\Lambda$ is absorbed by the trap, whereas the component outside this subspace, i.e. in $\bar{\Lambda}=\mathcal{H}\setminus \Lambda$, remains in the graph (see Fig. \ref{fig:nonunitary_dyn}). Let us expand the initial state on the basis of the eigenstates of the Hamiltonian

\begin{equation}
\ket{\psi_0}=\sum_{k=1}^m \braket{\lambda_k}{\psi_0} \ket{\lambda_k}+\sum_{k=m+1}^N \braket{\lambda_k}{\psi_0} \ket{\lambda_k}=\ket{\psi_\Lambda}+\ket{\psi_{\bar{\Lambda}}}\,,
\label{eq:psi0_Lambda_cmpnts}
\end{equation}
where we assume the eigenstates form an orthonormal basis\footnote{In case of degenerate energy levels we consider the eigenstates after orthonormalization.} and are ordered in such a way that $\Lambda = \operatorname{span}(\lbrace\ket{\lambda_k} \mid 1 \leq k \leq m \rbrace)$ and $\bar{\Lambda} = \operatorname{span}(\lbrace\ket{\lambda_k} \mid m+1 \leq k \leq N \rbrace)$. Then, the components in $\bar{\Lambda}$ are not affected by the open-dynamics which acts at the trap vertex $w$. The remaining components evolve in the subspace $\Lambda$ defined by having a finite overlap with the trap and are therefore absorbed at the trap. In the limit of $t\to+\infty$ the net result is the following: the total survival probability of finding the walker in the graph is $\braket{\psi_{\bar{\Lambda}}}{\psi_{\bar{\Lambda}}}\leq 1$, i.e. it is due to the part of the initial state expansion in $\bar{\Lambda}$; instead, the part of the initial state expansion in $\Lambda$ is fully absorbed at the trap, and so $\eta = \braket{\psi_\Lambda}{\psi_\Lambda}=\sum_{k=1}^m \vert \braket{\lambda_k}{\psi_0}\vert^2$. A further consequence of this is that if the system is initially prepared in a state $\ket{\psi_0}\in \bar{\Lambda}$, then the walker will stay forever in the graph without reaching the trap ($\eta=0$); if the system is initially prepared in a state $\ket{\psi_0}\in \Lambda$, then the walker will be completely absorbed by the trap ($\eta=1$).

If on the one hand this analytical technique allows one to compute the transport efficiency without solving dynamical equations, on the other hand diagonalizing the Hamiltonian still might be a hard task. The dimensionality reduction method in Sec. \ref{sec:dimrecmeth} allows one to avoid diagonalizing the Hamiltonian, since it can be proved that $\Lambda(H,\ket{w})=\mathcal{I}(H,\ket{w})$ (see Appendix \ref{app:subspace_equality}). Hence, we compute the transport efficiency as

\begin{equation}
\eta = \sum_{k=1}^m \left\vert \braket{e_k}{\psi_0}\right\vert^2\,,
\label{eq:transport_eff_ideal}
\end{equation}
i.e. as the overlap of the initial state $\ket{\psi_0}$ with the subspace $\mathcal{I}(H,\ket{w})=\operatorname{span}(\lbrace \ket{e_k} \mid 1\leq k \leq m \rbrace)$.

We consider as the initial state either a state localized at a vertex, $\ket{\psi_0}=\ket{v}$, or a superposition of two vertices,
$\ket{\psi_0}=(\ket{v_1}+e^{i\theta}\ket{v_2})/\sqrt{2}$. The localized initial state is a paradigmatic choice to take into account the fact that an excitation is usually created locally in a system. We also considered a superposition to investigate possible effects of coherence. The transport efficiency for the superposition of two vertices 

\begin{equation}
\eta_s = \frac{1}{2}\sum_{k=1}^m \left\vert \braket{e_k}{v_1}+e^{i\theta}\braket{e_k}{v_2}\right\vert^2
\label{eq:transport_eff_super}
\end{equation}
can be easily assessed, in some cases, when knowing the transport efficiency $\eta_{1}$ and $\eta_{2}$ for an initial state localized at $v_{1}$ and $v_2$, respectively. If $\ket{v_1}$ and $\ket{v_2}$ have the same overlap with the basis states, i.e. $\braket{e_k}{v_1}=\braket{e_k}{v_2}$ for $1\leq k \leq m$, then $\eta_1=\eta_2=\eta$ and we have

\begin{equation}
\eta_s(\theta)=\frac{1}{2}\left \vert 1+e^{i\theta}\right\vert^2 \eta =(1+\cos \theta)\eta\,,
\label{eq:eta_s_same_overlap}
\end{equation}
so $0\leq \eta_s(\theta) \leq 2\eta$. Instead, if $\ket{v_1}$ and $\ket{v_2}$ have nonzero overlap with different basis states, i.e. $\braket{e_k}{v_1}\neq0$ for $1\leq k \leq m_1$ and $\braket{e_k}{v_2}\neq0$ for $m_1+1\leq k \leq m_2$, with $m_2 \leq m$, then we have

\begin{equation}
\eta_s =\frac{1}{2}(\eta_1+\eta_2)\,,
\label{eq:eta_s_orthogonal_overlap}
\end{equation}
and it is does not depend on $\theta$.

In the following sections we study quantum transport on different graphs which are relevant in terms of symmetry, regularity, and connectivity. For each graph, we determine the basis of the subspace in which the system evolves, the reduced Hamiltonian \eqref{eq:qtH}, and the transport efficiency \eqref{eq:transport_eff_ideal} for an initial state localized at a vertex or a superposition of two vertices which is not covered by Eq. \eqref{eq:eta_s_same_overlap}. To analytically deal with a graph, we will group together the vertices which evolve identically by symmetry \cite{janmark2014global,wong2015diagrammatic,meyer2015connectivity,wong2016laplacian}. We mean that such vertices behave identically under the action of the Hamiltonian, in the sense that they are equivalent upon relabeling of vertices, as well as, e.g., all the vertices in a complete graph are equivalent. This does not mean that the time evolution $\ket{v_1(t)}$ of an initial state localized at a vertex $v_1$ is exactly equal to the time evolution $\ket{v_2(t)}$ of another initial state localized at $v_2 \neq v_1$, but it means that these two time evolutions are the same upon exchanging the labels of the two vertices. Note that the Hamiltonian \eqref{eq:qtH} acts on a generic vertex as the Laplacian, except for the trap vertex, which thus forms a subset of one element, itself. The equal superpositions of the vertices in each subset form a orthonormal basis for a subspace of the Hilbert space and the Hamiltonian written in such basis still fully describes the evolution of the system. However, we point out that such basis spans a subspace which, in general, is not the subspace $\mathcal{I}(H,\ket{w})$ we need to compute the transport efficiency. Nevertheless, this grouping of vertices provides a useful framework to analytically deal with the system, and for this reason we will introduce it. Clearly, identically evolving vertices have the same transport properties. However, vertices which are not equivalent for the Hamiltonian can provide the same transport efficiency. For this reason, in the following we will stress when this is the case.

\subsection{Complete bipartite graph}
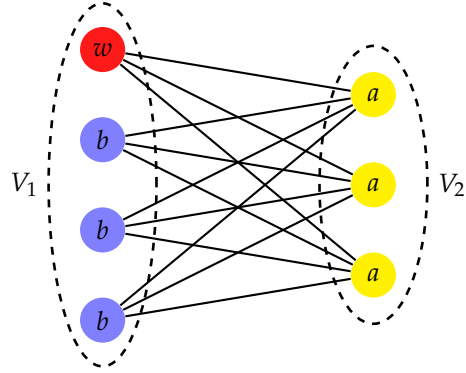
\begin{figure}[tb!]
\centering
\begin{tikzpicture}
[scale=1.2,every node/.style={circle,inner sep=0pt,minimum size=0.6cm}] %draw=black,
%Partition V
\node[fill=blue!50] (v1) at (-1.5,-1.5) {$b$}; 
\node[fill=blue!50] (v2) at (-1.5,-0.5) {$b$}; 
\node[fill=blue!50] (v3) at (-1.5,0.5) {$b$};  
\node[fill=red!90](v4) at (-1.5,1.5){$w$};
\node[ellipse, draw=black,fill=none,style=dashed,line width=1, fit=(v1) (v2) (v3) (v4), inner xsep=2mm, inner ysep=-4mm,label=left:$V_1$] (V1) {}; 
%Partition W
\node[fill=yellow] (w1) at (1.5,-1){$a$}; 
\node[fill=yellow] (w2) at (1.5,0) {$a$};
\node[fill=yellow] (w3) at (1.5,1) {$a$}; 
\node[ellipse, draw=black,fill=none,style=dashed,line width=1, fit=(w1) (w2) (w3), inner xsep=2mm, inner ysep=-2mm,label=right:$V_2$] (V2) {};
% Edges
\foreach \from in {v1,v2,v3,v4}
	\foreach \to in {w1,w2,w3}
    \draw[style={thick}] (\from) -- (\to);   
\end{tikzpicture}
\caption{Complete bipartite graph $K_{4,3}$. The trap vertex $w\in V_1$ is colored red. Identically evolving vertices have same transport properties and are identically colored and labeled.}
\label{fig:CBG}
\end{figure}

The complete bipartite graph (CBG) $G(V_1,V_2, E)$ is a highly symmetrical structure which, in general, is not regular. The CBG has two sets of vertices, $V_1$ and $V_2$, such that each vertex of $V_1$ is only connected to all the vertices of $V_2$ and vice versa. The set of CBGs is usually denoted as $K_{N_1,N_2}$, where the orders of the two partitions $N_1=\vert V_1 \vert$ and $N_2=\vert V_2 \vert$ are such that $N_1+N_2=N$, with $N$ the total number of vertices. The CBG is non regular as long as $N_1 \neq N_2$ (see $K_{4,3}$ in Fig. \ref{fig:CBG}), and the star graph is a particular case of CBG with $N_1=N-1$ and $N_2=1$. Without loss of generality, we assume the trap vertex $w\in V_1$.

The system evolves in a $3$-dimensional subspace (see Appendix \ref{app:basis_cbg}) spanned by the orthonormal basis states

\begin{equation}
\ket{e_1}=\ket{w}\,, \quad 
\ket{e_2} =\frac{1}{\sqrt{N_2}}\sum_{i \in V_2}\vert i \rangle\,, \quad 
\ket{e_3}=\frac{1}{\sqrt{N_1-1}}\sum_{\substack{i \in V_1,\\i\neq w}}\vert i \rangle\,.
\label{eq:basis_cbg}
\end{equation}
%In Fig. \ref{fig:CBG}, the state $\ket{e_2}$ is the equal superposition of the yellow $a$ vertices, and $\ket{e_3}$ that of the blue $b$ vertices.
This is also the basis we would obtain by grouping together the identically evolving vertices in the subsets $V_a=V_2$ and $V_b=V_1 \setminus \lbrace w \rbrace$ (see Fig. \ref{fig:CBG}) \cite{wong2016laplacian}. In this subspace the reduced Hamiltonian is

\begin{equation}
H=\begin{pmatrix}
(1-\alpha) N-i\kappa & -\sqrt{(1-\alpha)N} & 0\\
-\sqrt{(1-\alpha)N} & \alpha N & -\sqrt{(1-\alpha)(\alpha N-1)N}\\
0 &  -\sqrt{(1-\alpha)(\alpha N-1)N} & (1-\alpha)N
\end{pmatrix}\,,
\label{eq:qtHred_cbg}
\end{equation}
where $\alpha= N_1/N\in \mathbb{Q}^+$, $N_2=(1-\alpha)N$, since $N_1+N_2=N$. Notice that for $G$ to be a CBG, $\alpha$ must satisfy the condition $1/N \leq \alpha \leq 1-1/N$.

\begin{figure}[tb]
\centering
\includegraphics[width=\textwidth]{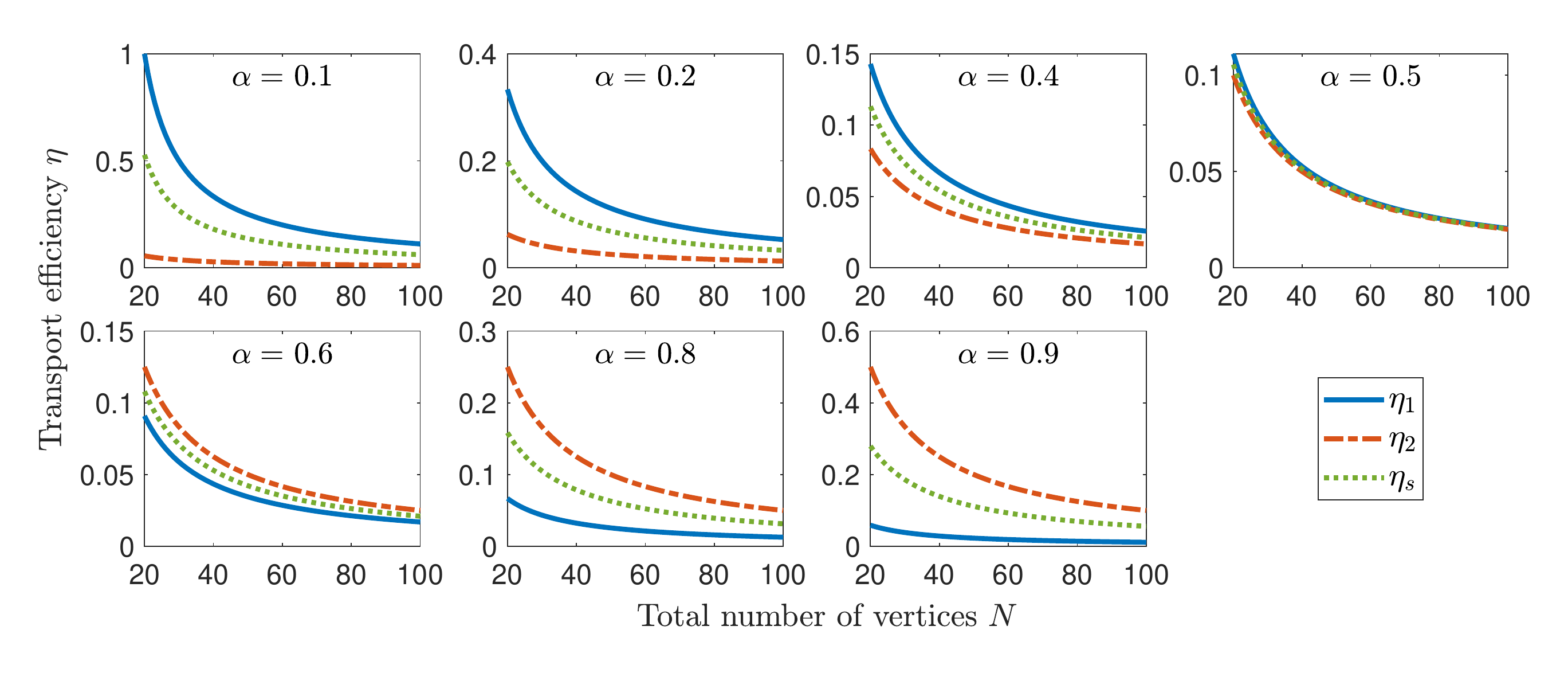}
\caption{Transport efficiency $\eta$ as a function of the order $N$ of the complete bipartite graph for different values of $\alpha=N_1/N$, with $N_1=\vert V_1\vert$, and different initial states. Transport efficiencies $\eta_{1(2)}$ \eqref{eq:eta_cbg_V1orV2} when the initial state is localized at a vertex in $V_{1(2)}$, and $\eta_s$ \eqref{eq:eta_cbg_s} when the initial state is the superposition of two vertices, one in $V_1$ and the other in $V_2$. The trap vertex $w \in V_1$.}
\label{fig:cbg_eta}
\end{figure}

If the initial state is localized at a vertex $v\neq w$, then the transport efficiency is

\begin{equation}
\eta = \begin{dcases}
\frac{1}{\alpha N-1} & \text{if $v \in V_1$ ,}\\
\frac{1}{(1-\alpha)N} & \text{if $v \in V_2$ ,}
\end{dcases}
\label{eq:eta_cbg_V1orV2}
\end{equation}
and we observe that

% \begin{equation}
% \eta_1 < \eta_2 \quad\Leftrightarrow\quad 2\alpha>1 \wedge N > \frac{1}{2\alpha-1}\,,
% \label{eq:eta_1less2}
% \end{equation}
\begin{equation}
\eta_1 < \eta_2 \quad\Leftrightarrow\quad 2\alpha>1+\frac{1}{N}\,,
\label{eq:eta_1less2}
\end{equation}
where $\eta_{1(2)}:=\eta(v \in V_{1(2)})$. Instead, if the initial state is a superposition of two vertices each of which belongs to a different partition, i.e. $v_1 \in V_1 \setminus \lbrace w \rbrace$ and $v_2\in V_2$, then the transport efficiency

\begin{equation}
\eta_{s} =\frac{N-1}{2N(\alpha N-1)(1-\alpha)}
\label{eq:eta_cbg_s}
\end{equation}
follows from Eq. \eqref{eq:eta_s_orthogonal_overlap}, so clearly $\eta_{2(1)} \leq \eta_s \leq \eta_{1(2)}$, where the alternative depends on the condition \eqref{eq:eta_1less2}. The transport efficiency depends on the parameters of the graph, $N$ and $\alpha$, as well as on the initial state (see Fig. \ref{fig:cbg_eta}). Whether we consider an initial localized state or a superposition of two localized states, the asymptotic behavior is $\eta=O(1/N)$ if both $N_1$ and $N_2$ are sufficiently large.

\subsection{Strongly regular graph}
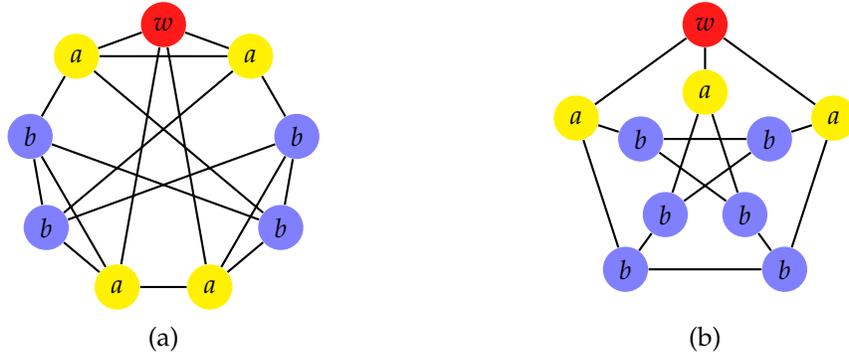
\begin{figure}[tb]
\centering
\begin{tikzpicture}
[scale=1.8,every node/.style={circle,inner sep=0pt,minimum size=0.6cm}]
\begin{scope}[shift={(-2,0)}]
\node[fill=red!90] (1) at ({50+1*360/9}:1) {$w$};
\node[fill=yellow] (2) at ({50+2*360/9}:1) {$a$};
\node[fill=blue!50] (3) at ({50+3*360/9}:1) {$b$};
\node[fill=blue!50] (4) at ({50+4*360/9}:1) {$b$};
\node[fill=yellow] (5) at ({50+5*360/9}:1) {$a$};
\node[fill=yellow] (6) at ({50+6*360/9}:1) {$a$};
\node[fill=blue!50] (7) at ({50+7*360/9}:1) {$b$};
\node[fill=blue!50] (8) at ({50+8*360/9}:1) {$b$};
\node[fill=yellow] (9) at ({50+0*360/9}:1) {$a$};
\draw[style={thick}]
(1)--(2)--(3)--(4)--(5)--(6)--(7)--(8)--(9)--(1)
(1)--(5)--(3)--(7)--(2)--(9)--(4)--(8)--(6)--(1)
;
\node[label=(a),minimum size=0cm] at (0,-1.5) {};
\end{scope}

\begin{scope}[shift={(2,0)}]
%Inner
\node[fill=yellow] (i1) at ({90+0*360/5}:0.5) {$a$};
\node[fill=blue!50] (i2) at ({90+1*360/5}:0.5) {$b$};
\node[fill=blue!50] (i3) at ({90+2*360/5}:0.5) {$b$};
\node[fill=blue!50] (i4) at ({90+3*360/5}:0.5) {$b$};
\node[fill=blue!50] (i5) at ({90+4*360/5}:0.5) {$b$};
%Outer
\node[fill=red!90] (o1) at ({90+0*360/5}:1) {$w$};
\node[fill=yellow] (o2) at ({90+1*360/5}:1) {$a$};
\node[fill=blue!50] (o3) at ({90+2*360/5}:1) {$b$};
\node[fill=blue!50] (o4) at ({90+3*360/5}:1) {$b$};
\node[fill=yellow] (o5) at ({90+4*360/5}:1) {$a$};
\draw[style={thick}]
(o1)--(o2)--(o3)--(o4)--(o5)--(o1);
\foreach \j in {1,2,...,5}{ \draw[style={thick}] (i\j)--(o\j);}
\draw[style={thick}]
(i4)--(i1)--(i3)--(i5)--(i2)--(i4);
\node[label=(b),minimum size=0cm] at (0,-1.5) {};
\end{scope}
\end{tikzpicture}
\caption{Two strongly regular graphs: (a) Paley graph with parameters $(9,4,1,2)$ (parametrization \eqref{eq:parameter_paley} for $\mu=2$); (b) Petersen graph with parameters $(10,3,0,1)$. The trap vertex $w$ is colored red. Identically evolving vertices have same transport properties and are identically colored and labeled.}
\label{fig:SRG}
\end{figure}

A strongly regular graph (SRG) with parameters $(N,k,\lambda,\mu)$ is a graph with $N$ vertices, not complete or edgeless, where each vertex is adjacent to $k$ vertices, for each pair of adjacent vertices there are $\lambda$ vertices adjacent to both, and for each pair of nonadjacent vertices there are $\mu$ vertices adjacent to both \cite{cameron1991designs,brouwer2011spectra}. If we consider the red vertex $w$ in Fig. \ref{fig:SRG}, this means that there are $k$ yellow adjacent vertices, and $N-k-1$ blue vertices, all at distance 2. SRGs have a local symmetry, but most have no global symmetry \cite{janmark2014global}. The four parameters $(N,k,\lambda,\mu)$ are not independent, and for some parameters there are no SRGs. One necessary but not sufficient condition is that the parameters satisfy

\begin{equation}
k(k-\lambda-1)=(N-k-1)\mu\,,
\label{eq:srg_condition}
\end{equation}
which can be proved by counting in two ways the vertices at distance $0$, $1$, and $2$ from a given vertex. Let us focus on the red vertex in Fig. \ref{fig:SRG} and count  the pairs of yellow and blue vertices adjacent to it. On the left-hand side of Eq. \eqref{eq:srg_condition}, the red vertex has $k$ neighbors, the yellow ones. Each yellow vertex has $k$ neighbors, one of which is the red one and $\lambda$ of which are other yellow vertices, so it is adjacent to $k-\lambda-1$ blue vertices. Hence, the number of pairs of adjacent yellow and blue vertices is $k(k-\lambda-1)$. On the right-hand side of Eq. \eqref{eq:srg_condition}, we consider the blue vertices, which, by definition, are not adjacent to the red vertex. There are $N-k-1$ blue vertices, since there are $N$ total vertices in the graph, one of which is red and $k$ of which are yellow. Each of the blue vertices is adjacent to $\mu$ yellow vertices, so there are $(N-k-1)\mu$ pairs of yellow and blue vertices. The condition \eqref{eq:srg_condition} comes from equating these expressions \cite{janmark2014global}.

The system evolves in a $3$-dimensional subspace (see Appendix \ref{app:basis_srg}) spanned by the orthonormal basis states

\begin{equation}
\ket{e_1}=\ket{w}\,, \quad 
\ket{e_2} =\frac{1}{\sqrt{k}}\sum_{(i,w) \in E}\vert i \rangle\,, \quad 
\ket{e_3}=\frac{1}{\sqrt{N-k-1}}\sum_{(i,w)\notin E}\vert i \rangle\,.
\label{eq:basis_srg}
\end{equation}
This is also the basis we would obtain by grouping together the identically evolving vertices in the subsets $V_a=\lbrace i \mid (i,w)\in E \rbrace$ and $V_b=\lbrace i \mid (i,w)\notin E\rbrace$ (see Fig. \ref{fig:SRG}) \cite{janmark2014global}. In this subspace the reduced Hamiltonian is

\begin{equation}
H=\begin{pmatrix}
k-i\kappa & -\sqrt{k} & 0\\
-\sqrt{k} & k-\lambda & -\sqrt{\mu(k-\lambda-1)}\\
0 & -\sqrt{\mu(k-\lambda-1)} & \mu
\end{pmatrix}\,.
\label{eq:qtHred_srg}
\end{equation}

If the initial state is localized at a vertex $v\neq w$, then the transport efficiency is

\begin{equation}
\eta =
\begin{dcases}
\frac{1}{k} & \text{if $(v,w) \in E$ ,}\\
\frac{1}{N-k-1} & \text{if $(v,w) \notin E$ .}
\end{dcases}
\label{eq:eta_srg_loc}
\end{equation}
Instead, if the initial state is a superposition of two vertices one of which is adjacent to $w$ and the other is not, i.e. $(v_1,w)\in E$ and $(v_2,w)\notin E$, then the transport efficiency

\begin{equation}
\eta_{s} = \frac{N-1}{2k(N-k-1)}
\label{eq:eta_srg_s}
\end{equation}
follows from Eq. \eqref{eq:eta_s_orthogonal_overlap}.

A family of SRGs is the Paley graphs (see Fig. \ref{fig:SRG}(a)), which are parametrized by

\begin{equation}
(N,k,\lambda,\mu)=(4\mu+1,2\mu,\mu-1,\mu)
\label{eq:parameter_paley}
\end{equation}
where $N$ must be a prime power\footnote{A prime power is a prime or integer power of a prime \cite{WolframPrimePower}.} such that $N\equiv 1\pmod{4}$. According to the parametrization \eqref{eq:parameter_paley}, whether we consider an initial localized state or a superposition of two localized states, the transport efficiency on a Paley graph is $\eta=1/2\mu$ (see Eqs. \eqref{eq:eta_srg_loc}--\eqref{eq:eta_srg_s}), regardless of the fact that the vertices considered are adjacent or not to $w$.

\subsection{Joined complete graphs}
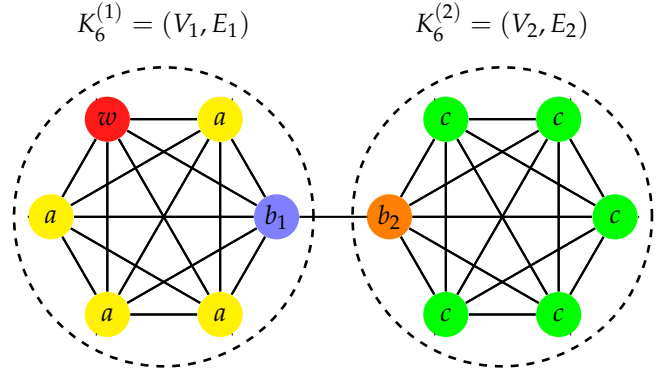
\begin{figure}[tb]
\centering
\begin{tikzpicture}
[scale=1.5,every node/.style={circle,inner sep=0pt,minimum size=0.6cm}]
\begin{scope}[shift={(-1.5,0)}]
		\node[fill=blue!50] (l1) at ({0*360/6}:1) {$b_1$};
		\node[fill=yellow] (l2) at ({1*360/6}:1) {$a$};
		\node[fill=red!90] (l3) at ({2*360/6}:1) {$w$};
		\node[fill=yellow] (l4) at ({3*360/6}:1) {$a$};
		\node[fill=yellow] (l5) at ({4*360/6}:1) {$a$};
		\node[fill=yellow] (l6) at ({5*360/6}:1) {$a$};
		
		\node[ellipse, draw=black,fill=none,style=dashed,line width=1, fit=(l1) (l2) (l3) (l4) (l5) (l6), inner xsep=-4mm, inner ysep=-2mm,label={[label distance=-6 mm]90:\text{$K_6^{(1)}=(V_1,E_1)$}}] (V1) {};		
		%Edges
		\foreach \from in {1,2,...,6}
			\foreach \to in {1,2,...,6}
	    		\draw[style={thick}] (l\from) -- (l\to);
\end{scope}	

\begin{scope}[shift={(1.5,0)}]
		\node[fill=green] (r1) at ({0*360/6}:1) {$c$};
		\node[fill=green] (r2) at ({1*360/6}:1) {$c$};
		\node[fill=green] (r3) at ({2*360/6}:1) {$c$};
		\node[fill=orange] (r4) at ({3*360/6}:1) {$b_2$};
		\node[fill=green] (r5) at ({4*360/6}:1) {$c$};
		\node[fill=green] (r6) at ({5*360/6}:1) {$c$};
		
		\node[ellipse, draw=black,fill=none,style=dashed,line width=1, fit=(r1) (r2) (r3) (r4) (r5) (r6), inner xsep=-4mm, inner ysep=-2mm,label={[label distance=-6 mm]90:\text{$K_6^{(2)}=(V_2,E_2)$}}] (V2) {};
		
		%Edges
		\foreach \from in {1,2,...,6}
			\foreach \to in {1,2,...,6}
	    		\draw[style={thick}] (r\from) -- (r\to);
\end{scope}	
%bridge
\draw[style={thick}] (l1) -- (r4);
\end{tikzpicture}
\caption{A graph with $12$ vertices constructed by joining two complete graphs of $6$ vertices by a single edge $(b_1,b_2)$, the bridge. The trap vertex $w \in V_1$ is colored red. Identically evolving vertices have same transport properties and are identically colored and labeled. The vertices $b_1$ and $b_2$ show the same transport efficiency even if they behave differently under the action of the Hamiltonian.}
\label{fig:JCG}
\end{figure}

The transport efficiency on a complete graph, when the initial state is localized at a vertex $v\neq w$, is $\eta=1/(N-1)$ \cite{caruso2009highly,novo2015systematic}. Here we consider two complete graphs of $N/2$ vertices joined by a single edge (see Fig. \ref{fig:JCG}). The two vertices, $b_1$ and $b_2$, forming the ``bridge'' have degree $N/2$, whereas all the others have degree $N/2-1$. We denote each complete graph by $K_{N/2}^{(k)}=(V_k,E_k)$, with $k=1,2$, where $\vert V_1 \vert=\vert V_2 \vert=N/2$. Then, the resulting joined graph is such that $V=V_1 \cup V_2$ and $E=E_1 \cup E_2 \cup \lbrace (b_1,b_2) \rbrace$. 

Grouping together the identically evolving vertices, we define the subsets $V_a=V_1 \setminus \lbrace w,b_1\rbrace$ and $V_c=V_2 \setminus \lbrace b_2\rbrace$ (see Fig. \ref{fig:JCG}). The system evolves in a $4$-dimensional subspace (see Appendix \ref{app:basis_jcg}) spanned by the orthonormal basis states

\begin{align}
\ket{e_1}&=\ket{w}\,,\nonumber\\
\ket{e_2}&=\frac{1}{\sqrt{N/2-1}}\left(\sum_{i \in V_a} \ket{i}+\ket{b_1}\right)\,, \nonumber\\
\ket{e_3}&=\frac{1}{\sqrt{(N-3)(N/2-1)}}\left[\sum_{i \in V_a}\ket{i}-(N/2-2)\ket{b_1}+(N/2-1)\ket{b_2}\right]\,,\nonumber\\
\ket{e_4}&=\frac{1}{\sqrt{(N-3)[N(N/2-2)+1]}}\left[\sum_{i \in V_a}\ket{i}-(N/2-2)(\ket{b_1}+\ket{b_2})-(N-3)\sum_{i \in V_c}\ket{i}\right]\,.
\label{eq:basis_jcg}
\end{align}
We point out that this basis spans a subspace of dimension $4$, thus smaller than the $5$-dimensional subspace spanned by the basis defined by grouping together the identically evolving vertices \cite{meyer2015connectivity}. In the subspace spanned by the basis states $\lbrace \ket{e_1},\ldots,\ket{e_4}\rbrace$ the reduced Hamiltonian is

\begin{equation}
\renewcommand{\arraystretch}{2.}
H=\begin{pmatrix}
N/2-1-i\kappa & -\sqrt{N/2-1} &0 &0\\
-\sqrt{N/2-1} & \frac{N}{N-2} &-\frac{\sqrt{N-3}}{N/2-1} &0\\
0 & -\frac{\sqrt{N-3}}{N/2-1}  &\frac{1}{N-3}\left(\frac{N^2}{2}-7+\frac{1}{N/2-1}\right) &\frac{\sqrt{(N/2-1)[N(N/2-2)+1]}}{N-3}\\
0 & 0 &\frac{\sqrt{(N/2-1)[N(N/2-2)+1]}}{N-3} &\frac{N/2-1}{N-3}
\end{pmatrix}\,.
\label{eq:qtHred_jcg}
\end{equation}

If the initial state is localized at a vertex $v\neq w$, then the transport efficiency is

\begin{equation}
\eta =
\begin{dcases}
\frac{2(N-1)}{N(N-4)+2} & \text{if $v \in V_a$ ,}\\
\frac{1}{2}+\frac{N-3}{N(N-4)+2} & \text{if $v \in \lbrace b_1,b_2\rbrace$ ,}\\
\frac{2(N-3)}{N(N-4)+2} & \text{if $v \in V_c$ .}
\end{dcases}
\label{eq:eta_jcg_loc}
\end{equation}
Assuming that each complete graph has $N/2 \geq 3$ vertices, then $\eta_c < \eta_a \leq \eta_b$, where the subscript refers to an initial state localized at vertex in $V_c$, in $V_a$, and in the bridge $\lbrace b_1,b_2\rbrace$, respectively. Instead, if the initial state is a superposition of two vertices, then

\begin{equation}
\eta_s(\theta) =
\begin{dcases}
\frac{(N-2)[N+4(1+\cos\theta)]}{4[N(N-4)+2]}=\frac{1}{4}+O\left(\frac{1}{N}\right)  & \text{if $v_1 \in V_a \wedge v_2 \in \lbrace b_1,b_2\rbrace$ ,}\\
\frac{2(N-2-\cos\theta)}{N(N-4)+2}=\frac{2}{N}+O\left(\frac{1}{N^2}\right) & \text{if $v_1 \in V_a \wedge v_2 \in V_c$ ,}\\
\frac{(N-2)[N-(N-4)\cos\theta]-4}{2[N(N-4)+2]}=\frac{1-\cos\theta}{2}+O\left(\frac{1}{N}\right)  & \text{if $v_1=b_1  \wedge v_2=b_2$ ,}\\
\frac{N(N+2)+4(N-4)\cos\theta-16}{4[N(N-4)+2]}=\frac{1}{4}+O\left(\frac{1}{N}\right)  & \text{if $v_1 \in \lbrace b_1,b_2\rbrace \wedge v_2 \in V_c$ .}
\end{dcases}\label{eq:eta_jcg_s}
\end{equation}

We observe that for the superposition of $v_1\in V_a$ and $v_2\in V_c$ the transport efficiency $\eta_s(\pi)$ is equal to $\eta$ for an initial state localized at $v \in V_a$. For the superposition of $b_1$ and $b_2$, i.e. of the vertices of the bridge, we have $\eta_s(\pi)=1$. This means that such state belongs to $\mathcal{I}(H,\ket{w})$, indeed

\begin{equation}
\frac{1}{\sqrt{2}}(\ket{b_1}-\ket{b_2})=\frac{1}{\sqrt{N-2}}(\ket{e_2}-\sqrt{N-3}\ket{e_3})\,.
\end{equation}
For an initial state localized at $b_1$ or $b_2$ we have the same transport efficiency $\eta_b$ \eqref{eq:eta_jcg_loc}. However, the two vertices $b_1$ and $b_2$ have different overlap with the basis states $\ket{e_k}$, so the transport efficiency \eqref{eq:eta_jcg_s} for the superposition of them is not given by Eq. \eqref{eq:eta_s_same_overlap}.

\subsection{Simplex of complete graphs}
\begin{figure}[tb]
\centering
\begin{tikzpicture}
[scale=0.8,every node/.style={circle,inner sep=0pt,minimum size=0.6cm}]
\begin{scope}[rotate=-120]
\foreach \n in {0,1,2,3}
	{
	\begin{scope}[shift={(\n*360/6:3)},rotate={90+\n*(60-360/5)}]
		\node[fill=cyan!20] (cg\n-0) at ({18+0*360/5}:1) {$f$};
		\node[fill=cyan!20] (cg\n-1) at ({18+1*360/5}:1) {$f$};
		\node[fill=magenta!70] (cg\n-2) at ({18+2*360/5}:1) {$e$};
		\node[fill=green] (cg\n-3) at ({18+3*360/5}:1) {$d$};
		\node[fill=cyan!20] (cg\n-4) at ({18+4*360/5}:1) {$f$};
		%Edges
		\foreach \from in {0,1,...,4}
			\foreach \to in {0,1,...,4}
	    		\draw[style={thick}] (cg\n-\from) -- (cg\n-\to);
	\end{scope}	
	}
\begin{scope}[shift={(4*360/6:3)},rotate={90+4*(60-360/5)}]
	\node[fill=orange] (cg4-0) at ({18+0*360/5}:1) {$c$};
	\node[fill=orange] (cg4-1) at ({18+1*360/5}:1) {$c$};
	\node[fill=orange] (cg4-2) at ({18+2*360/5}:1) {$c$};
	\node[fill=blue!50] (cg4-3) at ({18+3*360/5}:1) {$b$};
	\node[fill=orange] (cg4-4) at ({18+4*360/5}:1) {$c$};
	%Edges
	\foreach \from in {0,1,...,4}
		\foreach \to in {0,1,...,4}
	   		\draw[style={thick}] (cg4-\from) -- (cg4-\to);
\end{scope}
\begin{scope}[shift={(5*360/6:3)},rotate={90+5*(60-360/5)}]
	\node[fill=yellow] (cg5-0) at ({18+0*360/5}:1) {$a$};
	\node[fill=yellow] (cg5-1) at ({18+1*360/5}:1) {$a$};
	\node[fill=yellow] (cg5-2) at ({18+2*360/5}:1) {$a$};
	\node[fill=red!90] (cg5-3) at ({18+3*360/5}:1) {$w$};
	\node[fill=yellow] (cg5-4) at ({18+4*360/5}:1) {$a$};
	%Edges
	\foreach \from in {0,1,...,4}
		\foreach \to in {0,1,...,4}
	   		\draw[style={thick}] (cg5-\from) -- (cg5-\to);
\end{scope}
\end{scope}
%%\Edges
%crossing the origin
\foreach \g in {0,1,2} \draw[style={thick}] let \n1={int(\g+1)}, \n3={int(\g+3)},\n4={int(mod(int(\n3+1),5))} in (cg\g-\n1) -- (cg\n3-\n4); %"let" works with points \p1,\p2,.. or numbers \n1,\n2,..

%outer loop
\foreach \j in {0,1,...,4} \draw[style={thick}] let \n1={int(mod(int(\j+1),6))},\n2={int(mod(int(\j+4),5))} in (cg\j-\n2) edge[bend right] (cg\n1-\n2); 
\draw[style={thick}] (cg5-4) edge[bend right] (cg0-3); 

\draw[style={thick}] (cg0-0) edge[bend left] (cg2-4);
\draw[style={thick}] (cg1-1) edge[bend left] (cg3-0);
\draw[style={thick}] (cg2-2) edge[bend left] (cg4-1);\draw[style={thick}] (cg3-3) edge[bend left] (cg5-2);
\draw[style={thick}] (cg4-4) edge[bend left] (cg0-2);
\draw[style={thick}] (cg5-0) edge[bend left] (cg1-3);

\end{tikzpicture}
\caption{$5$-simplex of complete graphs. The trap vertex $w$ is colored red. Identically evolving vertices have same transport properties and are identically colored and labeled. The vertices in $V_c$ and $V_d$ show the same transport efficiency even if they behave differently under the action of the Hamiltonian.}
\label{fig:simplexCG}
\end{figure}
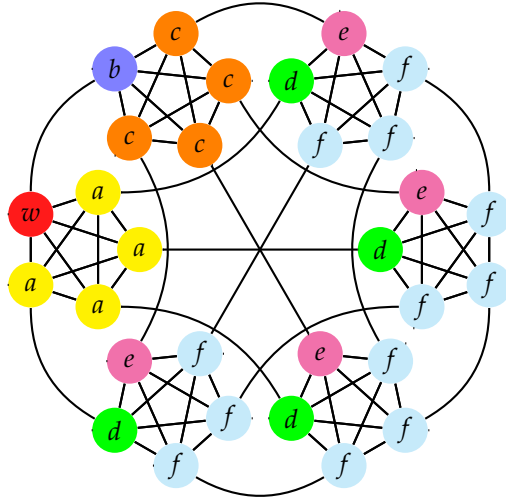

We call $M$-simplex of complete graphs what is formally known as the first-order truncated $M$-simplex lattice\footnote{The truncated $M$-simplex lattice is a generalization of the truncated tetrahedron lattice \cite{nelson1975soluble} and it is defined recursively. The graph of the zeroth order truncated $M$-simplex lattice is a complete graph of $M+1$ vertices. The graph for the $(n+1)$th order lattice is obtained by replacing each of the vertices of the $n$th order graph with a complete graph of $M$ vertices. The truncated simplex lattice has been studied in various problems, e.g. in statistical models \cite{dhar1977lattices}, self-avoiding random walks \cite{dhar1978self}, and spatial search \cite{meyer2015connectivity,wang2020optimal}.}. It is obtained by replacing each of the $M+1$ vertices of a complete graph with a complete graph of $M$ vertices (see Fig. \ref{fig:simplexCG}). Each of the new $M$ vertices is connected to one of the edges coming to the original vertex. The graph is regular, vertex transitive and there are $N=M(M+1)$ total vertices. 

Grouping together the identically evolving vertices, we define the subsets $V_a$, $V_c$, $V_d$, $V_e$, and $V_f$\footnote{The yellow vertices $a$ are adjacent to $w$ and belong to the same complete graph. The blue vertex $b$ is adjacent to $w$ but belongs to a different complete graph. The orange vertices $c$ are adjacent to $b$ and belong to the same complete graph. The green vertices $d$, even if at distance $2$ from $w$, like the vertices $c$, are adjacent to $a$, and so they form a different subset. The magenta vertices $e$ are adjacent to $c$ and belong to complete graphs other than the one the vertices $c$ belong to. The cyan vertices $f$ are adjacent to $e$ and $d$.} (see Fig. \ref{fig:simplexCG}), having cardinality $\vert V_a \vert =\vert V_c \vert =\vert V_d \vert =\vert V_e \vert =M-1$, and $\vert V_f \vert =(M-1)(M-2)$. Independently of $M$, the system evolves in a $5$-dimensional subspace (see Appendix \ref{app:basis_simplexcg}) spanned by the orthonormal basis states

\begin{align}
\ket{e_1}=&\ket{w}\,,\nonumber\\
\ket{e_2}=&\frac{1}{\sqrt{M}}\left( \sum_{i\in V_a} \ket{i}+\ket{b}\right)\,,\nonumber\\
\ket{e_3}=&\frac{\sqrt{M}}{\sqrt{(M-1)(M^2-2M+4)}}\left\lbrace\frac{M-2}{M} \left[\sum_{i\in V_a} \ket{i}-(M-1)\ket{b}\right]+\sum_{i\in V_c \cup V_d} \ket{i}\right\rbrace\,,\nonumber\\
\ket{e_4}=&\frac{\sqrt{M^2-2M+4}}{\sqrt{(M-1)(M^3+2M^2-8M+16)}}\left\lbrace\vphantom{\frac{(M)^2}{M^2}\sum_{i\in V_f}} \frac{2(M-2)}{M^2-2M+4} \left[ \sum_{i\in V_a} \ket{i}-(M-1)\ket{b}\right]\right.\nonumber\\
&\left.-\frac{(M-2)^2}{M^2-2M+4}\sum_{i\in V_c\cup V_d} \ket{i}-2\sum_{i\in V_e} \ket{i}-\sum_{i\in V_f} \ket{i}\right\rbrace\,,\nonumber\\
\ket{e_5}=&\frac{1}{M\sqrt{(M-1)(M-2)(M^3+2M^2-8M+16)}}\left\lbrace\vphantom{\sum_{i\in V_f}} -4(M-2) \left[ \sum_{i\in V_a} \ket{i}-(M-1)\ket{b}\right]\right.\nonumber\\
&\left.+2(M-2)^2\sum_{i\in V_c \cup V_d} \ket{i}-M^2(M-2)\sum_{i\in V_e} \ket{i}+2(M^2-2M+4)\sum_{i\in V_f} \ket{i}\right\rbrace\,.
\label{eq:basis_simplexCG}
\end{align}
Note that when the basis states include the vertices in $V_c$ and $V_d$, they always involve the equal superposition of all the vertices in $V_c\cup V_d$. Thus, these vertices are equivalent for quantum transport, even if they behave differently under the action of the Hamiltonian. We point out that this basis spans a subspace of dimension $5$, thus smaller than the $7$-dimensional subspace spanned by the basis defined by grouping together the identically evolving vertices \cite{wong2015diagrammatic,meyer2015connectivity}. In the subspace spanned by the basis states $\lbrace \ket{e_1},\ldots,\ket{e_5}\rbrace$ the reduced Hamiltonian is a symmetric tridiagonal matrix with cumbersome elements, so we store the main diagonal and the superdiagonal as follows

% \strut : to guarantee that an element on a page has a certain minimal height. 
\begin{equation}
\begin{pmatrix}
H_{1,1} & H_{1,2}\\
\vdots &\vdots\\
H_{n,n} & H_{n,n+1}\\
\vdots &\vdots\\
H_{5,5} & \ast
\end{pmatrix}
=
\begin{pmatrix}
M-i\kappa & -\sqrt{M}\\
\frac{\strut 3M-2}{\strut M} &-\frac{\strut \sqrt{(M-1)(M^2-2M+4)}}{\strut M}\\
\frac{\strut M^4- 2 M^3 + 4 M^2  - 4 M + 8 }{\strut M (M^2- 2 M +4)} & \frac{\strut \sqrt{M(M^3+2M^2-8M+16)}}{\strut M^2-2M+4}\\
\frac{\strut M(M^4-2M^3+20M^2-40M+64)}{\strut (M^3+2M^2-8M+16)(M^2- 2 M +4)} & \frac{\strut M(M+2)\sqrt{(M-2)(M^2-2M+4)}}{\strut M^3+2M^2-8M+16}\\
\frac{\strut (M+2)(M^3-4M+8)}{\strut M^3+2M^2-8M+16} & \ast
\end{pmatrix}\,,
\label{eq:Hred_simplexCG}
\end{equation}
where the $\ast$ denotes the missing element because its index exceeds the size of the matrix.

\begin{figure}[tb]
\centering
\includegraphics[width=0.7\textwidth]{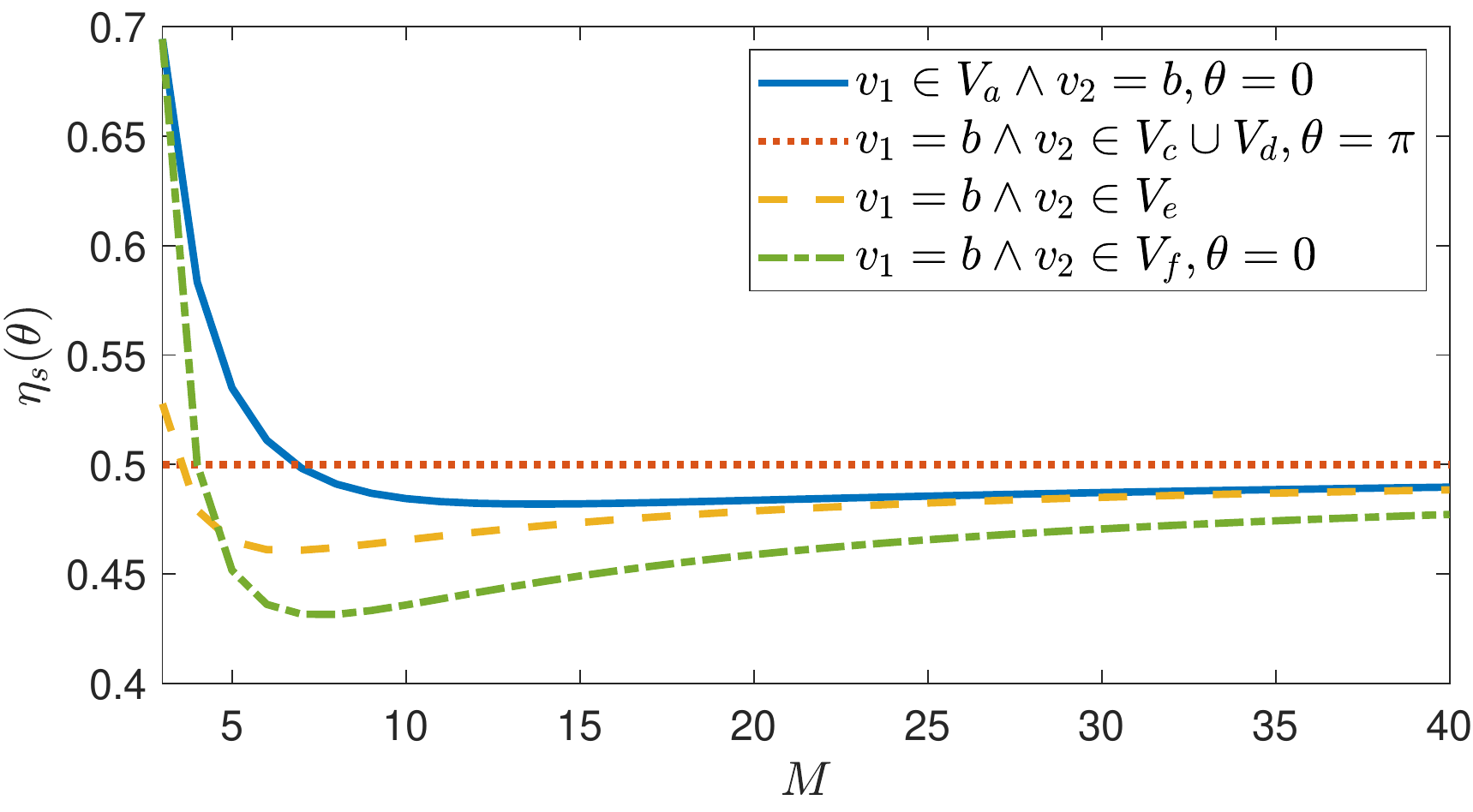}
\caption{Transport efficiency $\eta_s(\theta)$ \eqref{eq:eta_simplexcg_s} as a function of $M$ for different initial states $\ket{\psi_0}=(\ket{v_1}+e^{i\theta}\ket{v_2})/\sqrt{2}$. $M$ is the number of vertices in each of the $M+1$ complete graphs forming the $M$-simplex. The initial states are the possible equal superposition of two vertices one of which is $b$.}
\label{fig:simplexSB_eta}
\end{figure}

If the initial state is localized at a vertex $v\neq w$, then the transport efficiency is

\begin{equation}
\eta =
\begin{dcases}
\frac{M^2-2}{M^2(M-1)} & \text{if $v \in V_a $ ,}\\
\frac{M^2-2M+2}{M^2} & \text{if $v = b $ ,}\\
\frac{2}{M^2} & \text{if $v \in V_c \cup V_d$ ,}\\
\frac{1}{M-1} & \text{if $v \in V_e$ ,}\\
\frac{M^2-2M+4}{M^2(M-1)(M-2)} & \text{if $v \in V_f$ .}
\end{dcases}
\label{eq:eta_simplexcg_loc}
\end{equation}
Note that for an initial state localized at $b$, which is the only vertex adjacent to $w$ which does not belong to the complete graph of $w$ (see Fig. \ref{fig:simplexCG}), we have $\eta_b \approx 1$ for large $M$. Instead, if the initial state is a superposition of two vertices, then

\begin{equation}
\eta_s(\theta) =
\begin{dcases}
\frac{M(M^2-2M+4)-4+4(M-1)\cos\theta}{2M^2(M-1)}=\frac{1}{2}+O\left(\frac{1}{M}\right)& \text{if $v_1 \in V_a \wedge v_2=b$ ,}\\
\frac{M^2+2M-4+2(M-2)\cos\theta}{2M^2(M-1)}=\frac{1}{2M}+O\left(\frac{1}{M^2}\right)& \text{if $v_1 \in V_a \wedge v_2 \in V_c \cup V_d$ ,}\\
\frac{1}{M}+\frac{1}{M^2}& \text{if $v_1 \in V_a \wedge v_2 \in V_e$ ,}\\
\frac{M(M^2-M-4)+8-4(M-2)\cos\theta}{2M^2(M-1)(M-2)}=\frac{1}{2M}+O\left(\frac{1}{M^2}\right)& \text{if $v_1 \in V_a \wedge v_2 \in V_f$ ,}\\
\frac{M^2-2M+4-2(M-2)\cos\theta}{2M^2}=\frac{1}{2}+O\left(\frac{1}{M}\right)& \text{if $v_1 = b \wedge v_2 \in V_c \cup V_d$ ,}\\
\frac{1}{M^2}-\frac{1}{M}+\frac{M}{2(M-1)}=\frac{1}{2}+O\left(\frac{1}{M}\right)& \text{if $v_1 = b \wedge v_2 \in V_e$ ,}\\
%\frac{M^3-2M^2+4M-2}{2M^2(M-1)}& \text{if $v_1 = b \wedge v_2 \in V_e$ ,}\\
\frac{M(M^3-5M^2+11M-12)+8}{2M^2(M-1)(M-2)}+\frac{2}{M^2}\cos\theta=\frac{1}{2}+O\left(\frac{1}{M}\right) & \text{if $v_1 = b \wedge v_2 \in V_f$ ,}\\
\frac{1}{M^2}+\frac{1}{2(M-1)}=\frac{1}{2M}+O\left(\frac{1}{M^2}\right)& \text{if $v_1 \in V_c \cup V_d \wedge v_2 \in V_e$ ,}\\
\frac{3M^2-8M+8+2(M-2)^2\cos\theta}{2M^2(M-1)(M-2)}=\frac{3/2+\cos\theta}{M^2}+O\left(\frac{1}{M^3}\right)& \text{if $v_1 \in V_c \cup V_d \wedge v_2 \in V_f$ ,}\\
\frac{1}{M^2}+\frac{1}{M}-\frac{1}{M-1}+\frac{1}{2(M-2)}=\frac{1}{2M}+O\left(\frac{1}{M^2}\right)& \text{if $v_1 \in V_e \wedge v_2 \in V_f$ .}
%\frac{M^3-M^2-2M+4}{2M^2(M-1)(M-2)}& \text{if $v_1 \in V_e \wedge v_2 \in V_f$ .}
\end{dcases}
\label{eq:eta_simplexcg_s}
\end{equation}
Whenever the superposition of two vertices involves the vertex $b$, we have $\eta_s \approx 1/2$ for large $M$, and, in particular, $\eta_s(\pi)=1/2$ for $v_1 = b \wedge v_2 \in V_c \cup V_d$, independently of $M$ (see Fig. \ref{fig:simplexSB_eta}). Whenever the superposition involves a vertex in $V_e$, the transport efficiency does not depend on $\theta$. Moreover, we observe that the equal superposition of the vertices in $V_e$ belongs to $\mathcal{I}(H,\ket{w}$, since

\begin{equation}
\frac{1}{\sqrt{M-1}}\sum_{i\in V_e} \ket{i}= -\frac{1}{\sqrt{M^3+2M^2-8M+16}}\left( 2\sqrt{M^2-2M+4}\ket{e_4}+M\sqrt{M-2}\ket{e_5}\right)\,,
\end{equation} 
and so this state provides $\eta =1$.

In the $M$-simplex of complete graphs the total number vertices is $N=M(M+1)$, so the asymptotic behavior of the transport efficiency must be understood according to $M=O(\sqrt{N})$.

\section{Measures of connectivity}
\label{sec:connectivity}
The vertex connectivity $v(G)$ and the edge connectivity $e(g)$ of a graph $G$ are, respectively, the number of vertices or edges we must remove to make $G$ disconnected \cite{fiedler1989laplacian}. These are the two most common measures of graph connectivity, and

\begin{equation}
v(G) \leq e(G) \leq \delta (G)\,,
\label{eq:conn_rel_ved}
\end{equation}
i.e. both $v(G)$ and $e(G)$ are upper bounded by the minimum degree of the graph $\delta(G)$ \cite{west2001introduction}.

Another measure follows from the Laplace spectrum of the graph. The second-smallest eigenvalue $a(G)$ of the Laplacian of a graph $G$ with $N\geq 2$ vertices is the algebraic connectivity \cite{fiedler1973algebraic,de2007old} and, to a certain extent, it is a good parameter to measure how well a graph is connected. In spectral graph theory it is well known, e.g., that a graph is connected if and only if its algebraic connectivity is different from zero. Indeed, the multiplicity of the Laplace eigenvalue zero of an undirected graph $G$ is equal to the number of connected components of $G$ \cite{brouwer2011spectra}. For a complete graph we know that $v(K_N)=e(K_N)=N-1$ and $a(K_N)=N$. Instead, for a noncomplete graph $G$ we have $a(G)\leq v(G)$, and so $a(G)\leq e(G)$ \cite{fiedler1989laplacian}. 

Results of the different measures of connectivity for each graph are shown in Table \ref{tab:connectivity}. Vertex, edge, and algebraic connectivities for the complete and the complete bipartite graphs are from \cite{fiedler1989laplacian}. The measures of connectivity for the $M$-simplex of complete graphs are from \cite{meyer2015connectivity}.

The vertex connectivity of a SRG is $v(G)=k$ \cite{brouwer2011spectra} and the edge connectivity is $e(G)=k$. The latter follows from Eq. \eqref{eq:conn_rel_ved}, since $\delta(G)=k$, or using the fact that if a graph has diameter $2$, as the SRG has \cite{beineke2004topics}, then $e(G)=\delta(G)$ \cite{west2001introduction}. To assess the algebraic connectivity, we need the Laplace spectrum. The eigenvalues of the adjacency matrix $A$ are 

\begin{equation}
\frac{1}{2}\left[\lambda - \mu \pm \sqrt{(\lambda-\mu)^2+4(k-\mu)}\right], \quad k\,,
\end{equation}
and the scaling of them with $N$ depends on the type of SRG. Indeed, SRGs can be classified into two types  \cite{cameron1991designs,west2001introduction,beineke2004topics}. Type I graphs, for which $(N-1)(\mu-\lambda)=2k$. This implies that $\lambda=\mu-1$, $k=2\mu$, and $N=4\mu+1$. They exists if and only if $N$ is the sum of two squares. Examples include the Paley graphs (see parametrization \eqref{eq:parameter_paley}). Type II graphs, for which $(\mu-\lambda)^2+4(k-\mu)$ is a perfect square $d^2$, where $d$ divides $(N-1)(\mu-\lambda)-2k$, and the quotient is congruent to $N-1 \pmod{2}$. Type I graphs are also type II graphs if and only if $N$ is a square \cite{cameron1991designs}. The Paley graph $(9,4,1,2)$ is an example of this (see Fig. \ref{fig:SRG}(a)). Not all the SRGs of type II are known, only certain parameter families, e.g. the Latin square graphs \cite{cameron1991designs}, and certain graphs, e.g. the Petersen graph (see Fig. \ref{fig:SRG}(b)), are. Hence, we consider the algebraic connectivity only for the SRGs of type I. According to the parametrization of the SRG of type I and to the fact that $D=kI$, the eigenvalues of $L=D-A$ are 

\begin{equation}
0, \quad \frac{1}{2}(N\mp \sqrt{N})\,,
\end{equation}
from which the algebraic connectivity is $a(G)=(N- \sqrt{N})/2$, since $\mu=(N-1)/4$ and $k=(N-1)/2$.

\begin{table}[tb]
\caption{The minimum degrees and vertex, edge, and algebraic connectivities of the graphs with $N$ vertices considered in this work. For these graphs, the vertex and the edge connectivities are equal. Note that in the $M$-simplex of complete graphs $N=M(M+1)$.}
\centering
\setlength{\tabcolsep}{16pt} % Default value: 6pt
\renewcommand{\arraystretch}{1.5} % Default value: 1
%% \tablesize{} %% You can specify the fontsize here, e.g., \tablesize{\footnotesize}. If commented out \small will be used.
\begin{tabular}{cccc}
\toprule
	\textbf{Graph} \boldmath{$G$ }& \boldmath{$\delta(G)$} & \boldmath{$v(G)=e(G)$ }& \boldmath{$a(G)$}\\\midrule
	Complete $K_N$ & $N-1$ & $N-1$ & $N$ \\
	Complete bipartite $K_{N_1,N_2}$ & $\min(N_1,N_2)$ & $\min(N_1,N_2)$ & $\min(N_1,N_2)$\\
	Strongly regular (Type I)& $(N-1)/2$ & $(N-1)/2$ & $(N- \sqrt{N})/2$\\
	Joined complete $K_{N/2}$ & $N/2-1$ & $1$ & $O(1/N)$ \\
	$M$-simplex & $M=O(\sqrt{N})$ & $M=O(\sqrt{N})$ & $1$ \\
\bottomrule
\end{tabular}
\label{tab:connectivity}
\end{table}

For the joined complete graphs we have $v(G)=e(G)=1$, because of the bridge (see Fig. \ref{fig:JCG}) \cite{chartrand2012first}. The Laplace spectrum is

\begin{equation}
0,\frac{N}{2},\frac{1}{4}\left[N+4\pm \sqrt{N(N+8)-16} \right]\,,
\end{equation}
from which the algebraic connectivity is $a(G)=[N+4- \sqrt{N(N+8)-16}]/4$.

Then, we assess whether connectivity of the graph may provide or not some bounds on the transport efficiency for an initial state localized at a vertex. First, we focus on the regular graphs considered in this work, for which $\delta(G)=v(G)=e(G)$ and this is equal to the degree. For a complete graph we have $1/a(G) \leq \eta = 1/(N-1)$, and $1/(N-1)$ is also the reciprocal of the degree. For a SRG of type I we have $\eta = 2/(N-1)\leq 1/a(G)$ for $\mu\geq 1$, and $2/(N-1)$ is also the reciprocal of the degree. Hence, from these two examples, we see that the reciprocal of the algebraic connectivity does not provide a common bound on $\eta$. For the $M$-simplex of complete graphs, we observe that $a(G)=1$, from whose reciprocal we obtain the obvious upper bound $\eta \leq 1$. Note also that, in general, the transport efficiency for an initial state localized at vertex of a regular graph is not the reciprocal of the degree, as shown, e.g., by the transport efficiency on a general SRG \eqref{eq:eta_srg_loc} (degree $k$) and on the $M$-simplex \eqref{eq:eta_simplexcg_loc} (degree $M$).

Now, we focus on the non-regular graphs. For the joined complete graphs the reciprocal of the vertex and edge connectivity provides the obvious bound $\eta \leq 1$, whereas neither the reciprocal of $\delta (G)$ nor that of $a(G)$ provide a unique bound on $\eta$. Indeed, they are an upper or lower bound on $\eta$ depending on the initial state and the order of the graph (see Eq. \eqref{eq:eta_jcg_loc}). For the CBG, the vertex, edge, and algebraic connectivity is $\min(N_1,N_2)$ and its reciprocal is an upper or lower bound on the transport efficiency \eqref{eq:eta_cbg_V1orV2} depending on the geometry of the graph. Indeed, we have $\eta_1 \leq \eta_2 \leq 1/\min(N_1,N_2)$ for $\alpha > 1/2$, i.e. $N_1> N_2$, and $1/\min(N_1,N_2) = \eta_2 \leq \eta_1$ for $\alpha \leq 1/2$, i.e. $N_1 \leq N_2$. 

\begin{figure}[tb]
\centering
\includegraphics[width=\textwidth]{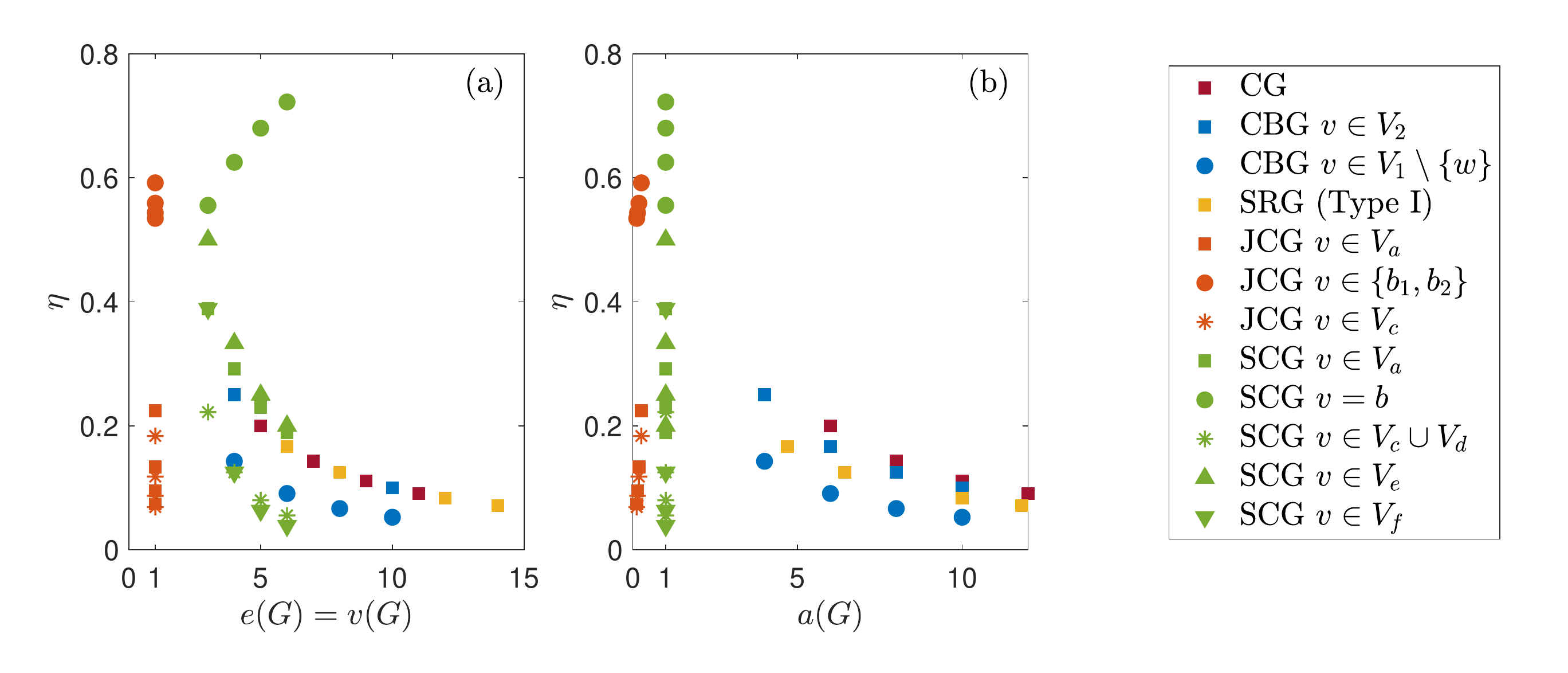}
\caption{Scatter plot of the correlation between the transport efficiency $\eta$ and (a) the edge or vertex connectivity, $e(G)$ and $v(G)$ respectively, or (b) the algebraic connectivity $a(G)$ (see also Table \ref{tab:connectivity}). Same color denotes results for the same graph: complete graph (CG, $N=6,8,10,12$), complete bipartite graph (CBG, $N=12,18,24,30$, $\alpha=2/3$), strongly regular graphs of type I (SRG, $N=13,17,25,29$), joined complete graphs (JCG, $N=12,18,24,30$), and $M$-simplex of complete graphs (SCG, $M=3,4,5,6$). For a given a graph, different markers denote initial states localized at different vertices $v$. Note that for the SRG of type I $\eta=1/2\mu=2/(N-1)$ independently of the fact that $(v,w)\in E$ or $(v,w)\notin E$. We observe some specific correlations between the transport efficiency and the connectivity for a given graph, but globally, among different graphs, transport efficiency and connectivity are uncorrelated. }
\label{fig:correl_eta_conn}
\end{figure}

In conclusion, just by focusing on the transport efficiency for an initial state localized at a vertex we observe that the connectivity is a poor indicator for the transport efficiency. First, because it does not provide any general lower or upper bound for estimating the transport efficiency, and transport efficiency and connectivity are generally uncorrelated (see Fig. \ref{fig:correl_eta_conn}). Second, because transport efficiency strongly depends on the initial state, or rather, on the overlap of this with the subspace spanned by the eigenstates of the Hamiltonian having non-zero overlap with the trap vertex, as shown in Sec. \ref{sec:qt}. Note that, analogously, we have found no general correlation between the transport efficiency and the normalized algebraic connectivity, which is the second-smallest eigenvalue of the normalized Laplacian matrix $\mathcal{L}$ of elements $\mathcal{L}_{jk}=L_{jk}/\sqrt{\deg(j)\deg(k)}$ \cite{chung1997spectral}.

\section{Conclusions}
\label{sec:conclusion}
In this work we have addressed the coherent dynamics of transport processes on graphs in the framework of continuous-time quantum walks. We have considered graphs having different properties in terms of regularity, symmetry, and connectivity and we have modeled the loss processes via the absorbing of the wavefunction component at a single trap vertex $w$. We have adopted the transport efficiency as a figure of merit to assess the transport properties of the system. In the ideal regime, as the one we have adopted, where there is no disorder nor decoherence processes during the transport, the transport efficiency $\eta$ can be computed as the overlap of the initial state with the subspace $\Lambda(H,\ket{w})$ spanned by the eigenstates of the Hamiltonian having non-zero overlap with the trap vertex. According to the dimensionality reduction method, we have determined the orthonormal basis of such subspace with no need to diagonalize the Hamiltonian. Therefore, any initial state which is a linear combination of such basis states provides the maximum transport efficiency $\eta =1$. We have considered as the initial state either a state localized at a vertex or a superposition of two vertices, and computed the corresponding transport efficiency. Overall, the most promising graph seems to be the $M$-simplex of complete graphs, since it allows us to have a transport efficiency close to $1$ for large $M$ for an initially localized state. Transport with maximum efficiency is also possible on other graphs, if the
walker is initially prepared in a suitable superposition state. However, the coherence of these preparations is likely to be degraded by noise, and the corresponding transport efficiency may be hard to be achieved in practice.

Our results suggest that connectivity of the graph is a poor indicator for the transport efficiency. Indeed, we observe some specific correlations between transport efficiency and connectivity for certain graphs, but in general they are uncorrelated. Moreover, transport efficiency depends on the overlap of the initial state with $\Lambda(H,\ket{w})$ and the reciprocal of the measures of connectivity we have assessed does not provide a general and consistent either lower or upper bound on $\eta$. However, the topology of the graph is encoded in the Laplacian matrix, which contributes to defining the Hamiltonian. Thus, connectivity somehow affects the transport properties of the system in the sense that it affects the Hamiltonian.

On the other hand, the transport efficiency is the integrated probability of trapping in the limit of infinite time, thus other figures of merit for the transport properties, such as the transfer time, which is the average time required by the walker to get absorbed at the trap, and the survival probability might highlight the role of the connectivity of the graph, if any. Moreover, the role of the trap needs to be further investigated, considering more than one trap vertex, different trapping rates, and different trap location. Our analytical results are proposed as a reference for further studies on the transport properties of these systems and as a benchmark for studying environment-assisted quantum transport on such graphs. Indeed, our work paves the way for further investigation including the analysis of more realistic systems in the presence of noise.

%\section{Discussion}
%Authors should discuss the results and how they can be interpreted in perspective of previous studies and of the working hypotheses. The findings and their implications should be discussed in the broadest context possible. Future research directions may also be highlighted.

%%%%%%%%%%%%%%%%%%%%%%%%%%%%%%%%%%%%%%%%%%

%\section{Conclusions}
%This section is not mandatory, but can be added to the manuscript if the discussion is unusually long or complex.

%%%%%%%%%%%%%%%%%%%%%%%%%%%%%%%%%%%%%%%%%%
%\section{Patents}
%This section is not mandatory, but may be added if there are patents resulting from the work reported in this manuscript.

%%%%%%%%%%%%%%%%%%%%%%%%%%%%%%%%%%%%%%%%%%
\vspace{6pt} 

%%%%%%%%%%%%%%%%%%%%%%%%%%%%%%%%%%%%%%%%%%
%% optional
%\supplementary{The following are available online at \linksupplementary{s1}, Figure S1: title, Table S1: title, Video S1: title.}


%%%%%%%%%%%%%%%%%%%%%%%%%%%%%%%%%%%%%%%%%%
\authorcontributions{Conceptualization, L.R., M.G.A.P. and P.B.; Methodology, L.R., M.G.A.P. and P.B.; Software, L.R.; Validation, L.R., M.G.A.P. and P.B.; Formal analysis, L.R., M.G.A.P. and P.B.; Investigation, L.R.; Writing--original draft preparation, L.R.; Writing--review and editing, L.R., M.G.A.P. and P.B.; Visualization, L.R.; Supervision, M.G.A.P. and P.B. All authors have read and agreed to the published version of the manuscript.}
%, please turn to the  \href{http://img.mdpi.org/data/contributor-role-instruction.pdf}{CRediT taxonomy} for the term explanation. Authorship must be limited to those who have contributed substantially to the work reported.}

%%%%%%%%%%%%%%%%%%%%%%%%%%%%%%%%%%%%%%%%%%
\funding{This research received no external funding.}
%\textcolor{red}{Please add: ``This research received no external funding'' or ``This research was funded by NAME OF FUNDER grant number XXX.'' and  and ``The APC was funded by XXX''. Check carefully that the details given are accurate and use the standard spelling of funding agency names at \url{https://search.crossref.org/funding}, any errors may affect your future funding.}

%%%%%%%%%%%%%%%%%%%%%%%%%%%%%%%%%%%%%%%%%%
\acknowledgments{P.B. and M.G.A.P. are members of GNFM-INdAM.}
%\textcolor{red}{In this section you can acknowledge any support given which is not covered by the author contribution or funding sections. This may include administrative and technical support, or donations in kind (e.g., materials used for experiments).}}

%%%%%%%%%%%%%%%%%%%%%%%%%%%%%%%%%%%%%%%%%%
\conflictsofinterest{The authors declare no conflict of interest.}
%Declare conflicts of interest or state ``The authors declare no conflict of interest.'' Authors must identify and declare any personal circumstances or interest that may be perceived as inappropriately influencing the representation or interpretation of reported research results. Any role of the funders in the design of the study; in the collection, analyses or interpretation of data; in the writing of the manuscript, or in the decision to publish the results must be declared in this section. If there is no role, please state ``The funders had no role in the design of the study; in the collection, analyses, or interpretation of data; in the writing of the manuscript, or in the decision to publish the results''.} 

%%%%%%%%%%%%%%%%%%%%%%%%%%%%%%%%%%%%%%%%%%
%% optional
\abbreviations{The following abbreviations are used in this manuscript:\\

\noindent 
\begin{tabular}{@{}ll}
CTQW & Continuous-time quantum walk\\
CBG & Complete bipartite graph\\
SRG & Strongly regular graph
\end{tabular}}

%%%%%%%%%%%%%%%%%%%%%%%%%%%%%%%%%%%%%%%%%%
%% optional
\appendixtitles{yes} % Leave argument "no" if all appendix headings stay EMPTY (then no dot is printed after "Appendix A"). If the appendix sections contain a heading then change the argument to "yes".
\appendix
\section{Subspace of the eigenstates of the Hamiltonian with non-zero overlap with the trap}
\label{app:subspace_equality}
In this appendix we show that the subspace $\Lambda(H,\ket{w})$ of the eigenstates of the Hamiltonian having nonzero overlap with the trap is equal to the subspace $\mathcal{I}(H,\ket{w})=\operatorname{span}(\lbrace H^k \ket{w} \mid k \in \mathbb{N}_0 \rbrace)$ introduced in Sec. \ref{sec:dimrecmeth}. This proof is from the \textit{Supplementary information} of \cite{novo2015systematic}. We report it for sake of completeness and because we refine a key point, not addressed in the original proof, about the right and the left inverse of a matrix.

Let $\Lambda(H,\ket{w})=\operatorname{span}(\lbrace \ket{\lambda_1},\ldots,\ket{\lambda_m}\rbrace)$, where $H\ket{\lambda_k}=\lambda_k \ket{\lambda_k}$ and $m$ is the minimum number of eigenstates of $H$ having non-zero overlap with the trap, i.e. $\braket{w}{\lambda_k}\neq 0$. In case of a degenerate eigenspace, more than one eigenstate belonging to it can have a non-zero overlap with $\ket{w}$, hence the need to find the minimum number $m$. The ambiguity is solved as follows. We choose the eigenstate from this degenerate eigenspace having the maximum overlap with $\ket{w}$, then we orthogonalize all the remaining eigenstates within such eigenspace with respect to it. After orthogonalizing, these eigenstates have zero overlap with $\ket{w}$ \cite{caruso2009highly,novo2015systematic}. 

Let $\dim(\mathcal{I}(H,\ket{w}))=m_1$, $\dim(\Lambda(H,\ket{w}))=m_2$, and $N$ the dimension of the complete Hilbert space. First, we prove that $\mathcal{I}(H,\ket{w}) \subseteq \Lambda (H,\ket{w})$, i.e. that any state $H^i \ket{w} \in \mathcal{I}(H,\ket{w})$ also belongs to $\Lambda (H,\ket{w})$:

\begin{equation}
H^i \ket{w}=\sum_{k=1}^N \braket{\lambda_k}{w}H^i\ket{\lambda_k}=\sum_{k=1}^{m_2} \braket{\lambda_k}{w}H^i\ket{\lambda_k}=\sum_{k=1}^{m_2} \braket{\lambda_k}{w}\lambda_k^i\ket{\lambda_k}\,,
\label{eq:Hw_in_Lambda} %%\textcolor
\end{equation}
since $\braket{\lambda_k}{w} = 0$ for $m_2+1\leq k \leq N$. Any state $H^i \ket{w}$ can therefore be expressed as a linear combination of the eigenstates of the Hamiltonian having a non-zero overlap with the trap, so $H^i \ket{w} \in \Lambda (H,\ket{w})\forall i \in \mathbb{N}_0$. Second, we prove that $\Lambda (H,\ket{w})\subseteq \mathcal{I}(H,\ket{w})$, i.e. that any state of $\Lambda (H,\ket{w})$ can be expressed as a linear combination of the states of $\mathcal{I}(H,\ket{w})$. We can write

\begin{align}
\ket{\lambda_j}=\sum_{i=1}^{m_1}c_{ji}H^{i-1}\ket{w}=\sum_{k=1}^{m_2}\sum_{i=1}^{m_1} c_{ji} \lambda_k^{i-1} \braket{\lambda_k}{w} \ket{\lambda_k}=\sum_{k=1}^{m_2} \sum_{i=1}^{m_1} c_{ji} M_{ik} \ket{\lambda_k}\,,
\label{eq:lambda_in_Hw} %%\textcolor
\end{align}
with matrix element $M_{ik}=\lambda_k^{i-1}\braket{\lambda_k}{w}$, provided that $\sum_{i=1}^{m_1} c_{ji} M_{ik} = \delta_{jk}$. In terms of matrices, this condition is $C_{m_2 \times m_1} M_{m_1 \times m_2} = I_{m_2 \times m_2}$, which means that $C$ is the left inverse of $M$, i.e. $C=M_L^{-1}$. Analogously, rewriting Eq. \eqref{eq:Hw_in_Lambda} and then using the first equality of Eq. \eqref{eq:lambda_in_Hw}, we have

\begin{equation}
H^{j-1} \ket{w}=\sum_{i=1}^{m_2} \braket{\lambda_i}{w}\lambda_i^{j-1}\ket{\lambda_i}=\sum_{i=1}^{m_2} M_{ji}\ket{\lambda_i}=\sum_{i=1}^{m_2}\sum_{k=1}^{m_1} M_{ji}c_{ik}H^{k-1}\ket{w}\,,
\end{equation}
provided that $\sum_{i=1}^{m_2} M_{ji}c_{ik} = \delta_{jk}$. In terms of matrices, this condition is $M_{m_1 \times m_2} C_{m_2 \times m_1}= I_{m_1 \times m_1}$, which means that $C$ is the right inverse of $M$, i.e. $C=M_R^{-1}$. Therefore, $M$ has a left and a right inverse, so $M$ must be square, $m_1=m_2=m$, and $M_L^{-1}=M_R^{-1}=M^{-1}=C$ is unique \cite{banerjee2014linear}. The condition under which $\Lambda (H,\ket{w})\subseteq \mathcal{I}(H,\ket{w})$ is thus that $M$ must be a $m\times m$ invertible matrix. The matrix $M$ is invertible if $\det(M)\neq 0$. We define two $m\times m$ matrices, $V_{ij}=\lambda_j^{i-1}$ and the diagonal matrix $D_{ij}=\delta_{ij}\braket{\lambda_j}{w}$, such that $M=VD$. Since $\braket{\lambda_j}{w}=0$ for $1\leq j \leq m$, then $\det(V)\neq0$. The matrix $V$ is of the Vandermonde form, so $\det(V)=\prod_{1\leq i < j \leq m}(\lambda_i-\lambda_j)$. This determinant is non-zero since all the states $\ket{\lambda_k}$, for $1\leq k \leq m$, belong to different eigenspaces, so all the $\lambda_k$ are different from each other. Hence, $\det(M)=\det(V)\det(D)\neq 0$, so $M$ is always invertible and this condition ensures that $\Lambda (H,\ket{w})\subseteq \mathcal{I}(H,\ket{w})$. This concludes the proof that $\Lambda (H,\ket{w}) = \mathcal{I}(H,\ket{w})$.

\section{Basis of $\mathcal{I}(H,\ket{w})$ for each graph}
\label{app:basis}
In this appendix we analytically derive the orthonormal basis $\lbrace \ket{e_k} \rbrace$ spanning the subspace $\mathcal{I}(H,\ket{w})$ for each graph considered. The first basis element is $\ket{e_1}=\ket{w}$, the trap vertex, and the $k$-th element $\ket{e_k}$ is obtained by orthonormalizing (O.N.) $H \ket{e_{k-1}}$ with respect to the subspace spanned by $\{\ket{e_1},\ldots,\ket{e_{k-1}}\}$. The procedure stops when we find the minimum $m$ such that $H\ket{e_m}\in \operatorname{span}(\{\ket{e_1},\ldots,\ket{e_m}\})$. The Hamiltonian \eqref{eq:qtH} is the sum of the Laplacian matrix, generating the CTQW on the graph, and the trapping Hamiltonian \eqref{eq:trapH}, which projects onto the trap $\ket{w}$ with proper coefficient.
\unskip
\subsection{Complete bipartite graph}
\label{app:basis_cbg}
The Laplacian matrix of the CBG $K_{N_1,N_2}$ is

\begin{equation}
L=N_2 \sum_{i \in V_1} \vert i \rangle \langle i \vert+N_1 \sum_{j \in V_2} \vert j \rangle \langle j \vert-\sum_{i \in V_1}\sum_{j \in V_2}(\vert i \rangle \langle j \vert+\vert j \rangle \langle i \vert)\,,
\end{equation}
since $\deg(i \in V_1)=N_2$ and $\deg(j \in V_2)=N_1$ (see Fig. \ref{fig:CBG}). The basis states \eqref{eq:basis_cbg} are obtained as follows:

\begin{align}
H\vert e_1 \rangle &= (N_2-i\kappa) \vert w \rangle -\sum_{j \in V_2} \vert j \rangle =(N_2-i\kappa)\vert e_1 \rangle-\sqrt{N_2}\vert e_2 \rangle \xrightarrow[]{\text{O.N.}} \vert e_2 \rangle\,,\\
H\vert e_2 \rangle &= \frac{N_1}{\sqrt{N_2}} \sum_{j \in V_2} \vert j \rangle -\frac{1}{\sqrt{N_2}}\sum_{i \in V_1}\sum_{j \in V_2} \vert i \rangle=  N_1 \vert e_2 \rangle-\sqrt{N_2}\sum_{\substack{i \in V_1,\\i\neq w}}\vert i \rangle - \sqrt{N_2} \vert e_1 \rangle\nonumber\\
&= N_1 \vert e_2 \rangle - \sqrt{N_2(N_1-1)}\vert e_3 \rangle-\sqrt{N_2}\vert e_1 \rangle\xrightarrow[]{\text{O.N.}} \vert e_3 \rangle\,,\\
H\vert e_3 \rangle &= \frac{N_2}{\sqrt{N_1-1}} \sum_{\substack{i\in V_1\\i\neq w}} \vert i \rangle -\frac{1}{\sqrt{N_1-1}}\sum_{\substack{i\in V_1\\i\neq w}}\sum_{j \in V_2} \vert j \rangle = N_2 \vert e_3 \rangle-\sqrt{N_2(N_1-1)} \vert e_2 \rangle \,.
\end{align}

In conclusion, any state $H^k \vert w \rangle \in \operatorname{span}(\left\lbrace \vert e_1 \rangle, \vert e_2 \rangle, \vert e_3 \rangle \right\rbrace) \forall k \in \mathbb{N}_0$, thus the states \eqref{eq:basis_cbg} form an orthonormal basis for the subspace $\mathcal{I}(H,\ket{w})$.

\subsection{Strongly regular graph}
\label{app:basis_srg}
The Laplacian matrix of the SRG with parameters $(N,k,\lambda,\mu)$ is

\begin{equation}
L=kI-\sum_{(j,i) \in E}\vert j \rangle \langle i \vert\,,
\end{equation}
where $I=\sum_{i \in V} \vert i \rangle \langle i \vert$ is the identity. Indeed, in a SRG each vertex has degree $k$, so the diagonal degree matrix is $D=kI$ (see Fig. \ref{fig:SRG}). The basis states \eqref{eq:basis_srg} are obtained as follows:

\begin{equation}
H\vert e_1 \rangle = (k-i\kappa)\vert e_1 \rangle-\sum_{(j,w) \in E} \vert j \rangle = (k-i\kappa) \vert e_1 \rangle- \sqrt{k}\vert e_2 \rangle \xrightarrow[]{\text{O.N.}} \vert e_2 \rangle\,.
\end{equation}

To address the computation of the next basis states, a remark is due. The diameter of a connected SRG $G$, i.e. the maximum distance between two vertices of $G$, is $2$ \cite{beineke2004topics}. This means that, given a vertex $w$, we can group all the other vertices in two subsets as follows: the subset of the vertices at a distance $1$ from $w$ (adjacent); the subset of the vertices at a distance $2$ from $w$ (nonadjacent). Because of the structure of the SRG, where two (non)adjacent vertices have $\lambda$ ($\mu$) common adjacent vertices, in the following we face summations with repeated terms.

To determine the third basis state we consider

\begin{equation}
H\vert e_2 \rangle = k \vert e_2 \rangle - \frac{1}{\sqrt{k}} \sum_{(i,w) \in E} \sum_{(j,i) \in E}\vert j \rangle
= (k-\lambda) \ket{e_2} - \sqrt{k}\ket{e_1}- \sqrt{\mu(k-\lambda-1)}\ket{e_3}\xrightarrow[]{\text{O.N.}} \vert e_3 \rangle\,.\label{eq:La_srg}
\end{equation}
To explain this, we have to focus on $\sum_{(i,w) \in E} \sum_{(j,i) \in E} \ket{j}$. The index of the first summation runs over the vertices $i$ adjacent to $w$, whereas the index of the second summation runs over the vertices $j$ adjacent to $i$. On the one hand, the vertex $w$ is counted $k$ times, because it has $k$ adjacent vertices $i$, each of which, in turn, has $j=w$ among its adjacent vertices. On the other hand, the index of the second summation runs over the vertices adjacent and nonadjacent to $w$, because of the structure of the SRG. Each vertex $j$ adjacent to $w$, i.e. $(j,w)\in E$, is connected to other $\lambda$ vertices adjacent to $w$, so it is counted $\lambda$ times.  Each vertex $j$ nonadjacent to $w$, i.e. $(j,w)\notin E$, is connected to $\mu$ vertices adjacent to $w$, so it is counted $\mu$ times. Thus we have

\begin{equation}
\sum_{(i,w) \in E} \sum_{(j,i) \in E} \ket{j}=k \ket{e_1}+\lambda \sum_{(j,w) \in E} \ket{j}+\mu\sum_{(j,w) \notin E} \ket{i}=k\ket{e_1}+\lambda\sqrt{k} \ket{e_2}+\mu \sqrt{N-k-1}\ket{e_3}\,.
\end{equation}
So, according to Eq. \eqref{eq:srg_condition}, we can write $\mu \sqrt{(N-k-1)}= \sqrt{\mu k(k-\lambda-1)}$, from which Eq. \eqref{eq:La_srg} follows.

Then, we consider 

\begin{equation}
H\ket{e_3}= k \ket{e_3} - \frac{1}{\sqrt{N-k-1}} \sum_{(i,w) \notin E} \sum_{(j,i) \in E}\ket{j} = \mu \ket{e_3}- \sqrt{\mu(k-\lambda-1)}\ket{e_2}\,. \label{eq:Lb_srg}
\end{equation}
Again, to explain this, we have to focus on the term $\sum_{(i,w) \notin E} \sum_{(j,i') \in E}\ket{j}$ in the second equality. The index of the first summation runs over the vertices $i$ nonadjacent to $w$, whereas the index of the second summation runs over the vertices $j$ adjacent to $i$. Each vertex $j$ nonadjacent to $w$, i.e. $(j,w)\notin E$, is connected to other $k-\mu$ vertices nonadjacent to $w$, so it is counted $k-\mu$ times.  Each vertex $j$ adjacent to $w$, i.e. $(j,w)\in E$, is connected to $k-\lambda-1$ vertices nonadjacent to $w$, so it is counted $k-\lambda-1$ times. Thus we have

\begin{align}
\sum_{(i,w) \notin E} \sum_{(j,i) \in E}\ket{j}&=(k-\lambda-1) \sum_{(i,w)\in E}\ket{i}+(k-\mu)\sum_{(i,w)\notin E}\ket{i}\nonumber\\
&=(k-\lambda-1)\sqrt{k}\ket{e_2}+(k-\mu)\sqrt{N-k-1}\ket{e_3}\,.
\end{align}
So, according to Eq. \eqref{eq:srg_condition}, we can write $(k-\lambda-1)\sqrt{k}= \sqrt{\mu (N-k-1)(k-\lambda-1)}$, from which Eq. \eqref{eq:Lb_srg} follows. 

In conclusion, any state $H^k \vert w \rangle \in \operatorname{span}(\left\lbrace \vert e_1 \rangle, \vert e_2 \rangle, \vert e_3 \rangle \right\rbrace) \forall k \in \mathbb{N}_0$, thus the states \eqref{eq:basis_srg} form an orthonormal basis for the subspace $\mathcal{I}(H,\ket{w})$.

\subsection{Joined complete graphs}
\label{app:basis_jcg}
The Laplacian matrix of the two complete graphs $K_{N/2}$ joined by a single edge $(b_1,b_2)$ is

\begin{equation}
L=L_1+L_2+\underbrace{\vert b_1 \rangle \langle b_1 \vert +\vert b_2 \rangle \langle b_2 \vert-\vert b_1 \rangle \langle b_2 \vert -\vert b_2 \rangle \langle b_1 \vert}_{\text{bridge}} \,,
\end{equation}
where 

\begin{equation}
L_k=\left(\frac{N}{2}-1\right)\sum_{i\in V_k} \vert i \rangle \langle i \vert - \sum_{(i,j)\in E_k} \vert i \rangle \langle j \vert
\end{equation}
is the Laplacian matrix of the complete graph $K_{N/2}^{(k)}$, with $k=1,2$. The bridge introduces the edge between the vertices $b_1$ and $b_2$ and correctly makes the degree of such vertices be $N/2$ (see Fig. \ref{fig:JCG}). Hence, $L \ket{v}=L_k \ket{v}$ for any vertex $v\in V_k \setminus \lbrace b_k\rbrace$. Instead, $L\ket{b_k}=(N/2)\ket{b_k}-\sum_{(i,b_k)\in E_k}\ket{i}-\ket{b_{\bar{k}}}$, where $\bar{k}$ is the complement of $k$ in $\lbrace 1,2\rbrace$. 

Reasoning by symmetry, we introduce the subsets of the identically evolving vertices, i.e. the subsets containing the vertices which behave identically under the action of the Hamiltonian:

\begin{align}
H\ket{w} &=(N/2-1-i\kappa)\ket{w}-\sum_{i \in V_a}\ket{i}-\ket{b_1}\,,\\
H\sum_{i \in V_a}\ket{i} &=2\sum_{i \in V_a}\ket{i}-(N/2-2)(\ket{w}+\ket{b_1})\,,\\
H\ket{b_1} &=N/2\ket{b_1} - \sum_{i \in V_a}\ket{i}-\ket{w}-\ket{b_2}\,,\\
H\ket{b_2} &=N/2\ket{b_2} - \sum_{i \in V_c}\ket{i}-\ket{b_1}\,,\\
H\sum_{i \in V_c}\ket{i} &=\sum_{i \in V_c}\ket{i}-(N/2-1)\ket{b_2}\,,
\end{align}
where $V_a=V_1 \setminus \lbrace w,b_1\rbrace$ and $V_c=V_2 \setminus \lbrace b_2\rbrace$. Note that the results of $H$ applied on the vertices $b_1$ or $b_2$ are different, and this is the reason why they form different subsets. According to these preliminary results, the basis states \eqref{eq:basis_jcg} are obtained as follows:

\begin{align}
H\ket{e_1}  =&(N/2-1-i\kappa)\ket{w}-\sum_{i \in V_a}\ket{i}-\ket{b_1} \xrightarrow[]{\text{O.N.}} \ket{e_2}\,,\\
H\ket{e_2}  =&\frac{1}{\sqrt{N/2-1}}\left[-(N/2-1) \ket{w}+\sum_{i \in V_a}\ket{i}+2 \ket{b_1}-\ket{b_2}\right]
\xrightarrow[]{\text{O.N.}} \ket{e_3}\,,\\
H\ket{e_3}  =& \frac{1}{\sqrt{(N-3)(N/2-1)}}\left[N/2\sum_{i \in V_a}\ket{i}-(N^2/4-3) \ket{b_1}+(N^2/4-2)\ket{b_2}\right.\nonumber\\
&\left.-(N/2-1)\sum_{i\in V_c}\ket{i} \right] \xrightarrow[]{\text{O.N.}} \ket{e_4}\,,
\end{align}
and it can be proved that

\begin{equation}
H\ket{e_4}=\frac{\sqrt{N/2-1}}{N-3}\left(\sqrt{N(N/2-2)+1}\ket{e_3}+\sqrt{N/2-1}\ket{e_4}\right)\,.
\end{equation}

In conclusion, any state $H^k \vert w \rangle \in \operatorname{span}(\left\lbrace \vert e_1 \rangle, \ldots, \vert e_4\rangle \right\rbrace) \forall k \in \mathbb{N}_0$, thus the states \eqref{eq:basis_jcg} form an orthonormal basis for the subspace $\mathcal{I}(H,\ket{w})$.

\subsection{Simplex of complete graphs}
\label{app:basis_simplexcg}
\begin{figure}[tb]
\centering
\begin{tikzpicture}[scale=0.6,every node/.style={circle,draw=none,inner sep=0pt,minimum size=0.4cm}]

\begin{scope}[shift={(-5,0)}]
\begin{scope}[shift={(0*360/6:3)},rotate={90+0*(60-360/5)}]
	\node[fill=magenta!70] (cg0-0) at ({18+0*360/5}:1) {$4$};
	\node[fill=green] (cg0-1) at ({18+1*360/5}:1) {$3$};
	\node[fill=none, draw=none,inner sep=0pt] at ({18+1*360/5}:-2) {$K_5^{(4)}$};
	\node[fill=cyan!20] (cg0-2) at ({18+2*360/5}:1) {$2$};
	\node[fill=cyan!20] (cg0-3) at ({18+3*360/5}:1) {$1$};
	\node[fill=cyan!20] (cg0-4) at ({18+4*360/5}:1) {$5$};
	%Edges
	\foreach \from in {0,1,...,4}
		\foreach \to in {0,1,...,4}
	   		\draw[style={thick}] (cg0-\from) -- (cg0-\to);
\end{scope}

\begin{scope}[shift={(1*360/6:3)},rotate={90+1*(60-360/5)}]
	\node[fill=magenta!70] (cg1-0) at ({18+0*360/5}:1) {$5$};
	\node[fill=green] (cg1-1) at ({18+1*360/5}:1) {$4$};
	\node[fill=cyan!20] (cg1-2) at ({18+2*360/5}:1) {$3$};
	\node[fill=none, draw=none,inner sep=0pt] at ({18+2*360/5}:-2) {$K_5^{(3)}$};
	\node[fill=cyan!20] (cg1-3) at ({18+3*360/5}:1) {$2$};
	\node[fill=cyan!20] (cg1-4) at ({18+4*360/5}:1) {$1$};
	%Edges
	\foreach \from in {0,1,...,4}
		\foreach \to in {0,1,...,4}
	   		\draw[style={thick}] (cg1-\from) -- (cg1-\to);
\end{scope}

\begin{scope}[shift={(2*360/6:3)},rotate={90+2*(60-360/5)}]
	\node[fill=orange] (cg2-0) at ({18+0*360/5}:1) {$1$};
	\node[fill=blue!50] (cg2-1) at ({18+1*360/5}:1) {$5$};
	\node[fill=orange] (cg2-2) at ({18+2*360/5}:1) {$4$};
	\node[fill=orange] (cg2-3) at ({18+3*360/5}:1) {$3$};
	\node[fill=none, draw=none,inner sep=0pt] at ({18+3*360/5}:-2) {$K_5^{(2)}$};
	\node[fill=orange] (cg2-4) at ({18+4*360/5}:1) {$2$};
	%Edges
	\foreach \from in {0,1,...,4}
		\foreach \to in {0,1,...,4}
	   		\draw[style={thick}] (cg2-\from) -- (cg2-\to);
\end{scope}

\begin{scope}[shift={(3*360/6:3)},rotate={90+3*(60-360/5)}]
	\node[fill=yellow] (cg3-0) at ({18+0*360/5}:1) {$2$};
	\node[fill=red!90] (cg3-1) at ({18+1*360/5}:1) {$1$};		
	\node[fill=yellow] (cg3-2) at ({18+2*360/5}:1) {$5$};
	\node[fill=yellow] (cg3-3) at ({18+3*360/5}:1) {$4$};
	\node[fill=yellow] (cg3-4) at ({18+4*360/5}:1) {$3$};
	\node[fill=none, draw=none,inner sep=0pt] at ({18+4*360/5}:-2) {$K_5^{(1)}$};
	%Edges
	\foreach \from in {0,1,...,4}
		\foreach \to in {0,1,...,4}
	   		\draw[style={thick}] (cg3-\from) -- (cg3-\to);
\end{scope}

\begin{scope}[shift={(4*360/6:3)},rotate={90+4*(60-360/5)}]
	\node[fill=cyan!20] (cg4-0) at ({18+0*360/5}:1) {$3$};
	\node[fill=none, draw=none,inner sep=0pt] at ({18+0*360/5}:-2) {$K_5^{(6)}$};
	\node[fill=magenta!70] (cg4-1) at ({18+1*360/5}:1) {$2$};
	\node[fill=green] (cg4-2) at ({18+2*360/5}:1) {$1$};
	\node[fill=cyan!20] (cg4-3) at ({18+3*360/5}:1) {$5$};
	\node[fill=cyan!20] (cg4-4) at ({18+4*360/5}:1) {$4$};
	%Edges
	\foreach \from in {0,1,...,4}
		\foreach \to in {0,1,...,4}
	   		\draw[style={thick}] (cg4-\from) -- (cg4-\to);
\end{scope}
\begin{scope}[shift={(5*360/6:3)},rotate={90+5*(60-360/5)}]
	\node[fill=cyan!20] (cg5-0) at ({18+0*360/5}:1) {$4$};
	\node[fill=magenta!70] (cg5-1) at ({18+1*360/5}:1) {$3$};
	\node[fill=none, draw=none,inner sep=0pt] at ({18+1*360/5}:-2) {$K_5^{(5)}$};
	\node[fill=green] (cg5-2) at ({18+2*360/5}:1) {$2$};
	\node[fill=cyan!20] (cg5-3) at ({18+3*360/5}:1) {$1$};
	\node[fill=cyan!20] (cg5-4) at ({18+4*360/5}:1) {$5$};
	%Edges
	\foreach \from in {0,1,...,4}
		\foreach \to in {0,1,...,4}
	   		\draw[style={thick}] (cg5-\from) -- (cg5-\to);
\end{scope}
%%\Edges
%crossing the origin
\foreach \g in {0,1,2} \draw[style={thick}] let \n1={int(\g+1)}, \n3={int(\g+3)},\n4={int(mod(int(\n3+1),5))} in (cg\g-\n1) -- (cg\n3-\n4); %"let" works with points \p1,\p2,.. or numbers \n1,\n2,..

%outer loop
\foreach \j in {0,1,...,4} \draw[style={thick}] let \n1={int(mod(int(\j+1),6))},\n2={int(mod(int(\j+4),5))} in (cg\j-\n2) edge[bend right] (cg\n1-\n2); 
\draw[style={thick}] (cg5-4) edge[bend right] (cg0-3); 

\draw[style={thick}] (cg0-0) edge[bend left] (cg2-4);
\draw[style={thick}] (cg1-1) edge[bend left] (cg3-0);
\draw[style={thick}] (cg2-2) edge[bend left] (cg4-1);\draw[style={thick}] (cg3-3) edge[bend left] (cg5-2);
\draw[style={thick}] (cg4-4) edge[bend left] (cg0-2);
\draw[style={thick}] (cg5-0) edge[bend left] (cg1-3);

\end{scope}

%BOXED LEGEND WITH NODES
\matrix [draw,rectangle,row sep=-8mm,column sep=6mm, inner xsep=2mm, inner ysep=-2mm,] at (5, 0) {
\node[fill=red, label=right:$w$] {}; & \node[fill=green, label=right:\text{$V_d=\lbrace d\rbrace$}] {};\\
\node[fill=yellow, label=right:\text{$V_a=\lbrace a\rbrace$}]  {}; &\node[fill=magenta!70, label=right:\text{$V_e=\lbrace e\rbrace$}]  {};\\
\node[fill=blue!50, label=right:$b$] {}; & \node[fill=cyan!20, label=right:\text{$V_f=\lbrace f\rbrace$}]  {};\\
\node[fill=orange, label=right:\text{$V_c=\lbrace c\rbrace$}]  {};\\
};
\end{tikzpicture}
\caption{Labeling of vertices in a $5$-simplex of complete graphs. The trap vertex $w$ is colored red and assumed to be $\ket{1}$ in $K_5^{(1)}$. Same coloring denotes the subsets $V_\alpha$ of identically evolving vertices $\alpha$, with $\alpha=w,a,b,c,d,e,f$ (see also Fig. \ref{fig:simplexCG}). Note that each of the two vertices $w$ and $b$ forms a subset of one element, itself.}
\label{fig:label_simplexCG}
\end{figure}
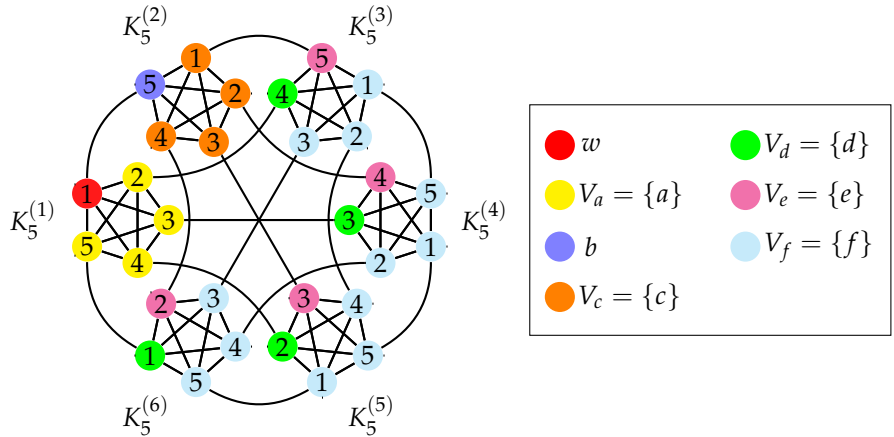

The Laplacian matrix is defined as $L=D-A$. For a $M$-simplex of complete graphs the diagonal degree matrix is $D=MI$, since the graph is regular, and the adjacency matrix is $A=\sum_{m=1}^{M+1} A_{\textup{intra}}^{(m)}+A_{\textup{inter}}$, where

\begin{equation}
A_{\textup{intra}}^{(m)}=\sum_{(i,j)\in E_m}\vert i^{(m)} \rangle \langle j^{(m)} \vert
\end{equation}
is the intra-graph adjacency matrix, i.e. within the complete graph $K_M^{(m)}$, and

\begin{equation}
A_{\textup{inter}}=\sum_{m=1}^{M+1}\sum_{i=1}^M \vert i^{(m)} \rangle \langle (M+1-i)^{(m')} \vert\,,
\label{eq:AinterSimplex}
\end{equation}
with $m'=1+\operatorname{mod}(i+m-1,M+1)$, is the inter-graphs adjacency matrix, i.e. between different complete graphs. The index $m$ labels the complete graphs $K_M^{(m)}$ forming the $M$-simplex. Note that Eq. \eqref{eq:AinterSimplex} follows the labeling of the vertices in Fig. \ref{fig:label_simplexCG} and it is just one of the possible ways to computationally implement the inter-graphs contribution.

In this case, using the notion of adjacency and reasoning by symmetry to introduce the subsets of the identically evolving vertices provide a framework which, analytically, is simpler and clearer to deal with than using explicitly the Laplacian above defined. These subsets contain the vertices which behave identically under the action of the Hamiltonian:

\begin{align}
H\ket{w} &=(M-i\kappa)\ket{w}-\sum_{i \in V_a}\ket{i}-\ket{b}\,,\\
H\sum_{i \in V_a}\ket{i} &=2\sum_{i \in V_a}\ket{i}-(M-1)\ket{w}-\sum_{i \in V_d}\ket{i}\,,\\
H\ket{b} &=M\ket{b} - \ket{w}-\sum_{i \in V_c}\ket{i}\,,\\
H\sum_{i \in V_c}\ket{i} &=2\sum_{i \in V_c}\ket{i}-(M-1)\ket{b}-\sum_{i \in V_e}\ket{i}\,,\\
H\sum_{i \in V_d}\ket{i} &=M\sum_{i \in V_d}\ket{i}-\sum_{i \in V_a}\ket{i}-\sum_{i \in V_e}\ket{i}-\sum_{i \in V_f}\ket{i}\,,\\
H\sum_{i \in V_e}\ket{i} &=M\sum_{i \in V_e}\ket{i}-\sum_{i \in V_c}\ket{i}-\sum_{i \in V_d}\ket{i}-\sum_{i \in V_f}\ket{i}\,,\\
H\sum_{i \in V_f}\ket{i} &=2\sum_{i \in V_f}\ket{i}-(M-2)\left(\sum_{i \in V_d}\ket{i}+\sum_{i \in V_e}\ket{i}\right)\,.
\end{align}
Note that the results of $H$ applied on the vertices in $V_c$ or in $V_d$ are different, and this is the reason why they form different subsets. According to these preliminary results, the basis states \eqref{eq:basis_simplexCG} are obtained as follows:

\begin{align}
H\ket{e_1} =&(M-i\kappa)\ket{w}-\sum_{i \in V_a}\ket{i}-\ket{b}\xrightarrow[]{\text{O.N.}} \ket{e_2}\,,\\
H\ket{e_2} =&\frac{1}{\sqrt{M}}\left( 2\sum_{i \in V_a}\ket{i}-M\ket{w}+M\ket{b}-\sum_{i \in V_c \cup V_d}\ket{i}\right)\xrightarrow[]{\text{O.N.}} \ket{e_3}\,,\\
H\ket{e_3} =&\frac{\sqrt{M}}{\sqrt{(M-1)(M^2-2M+4)}}\left[\vphantom{\sum_{i\in V_f}} \frac{M-4}{M}\sum_{i \in V_a}\ket{i}-(M-1)^2\ket{b}\right.\nonumber\\
&\left.+\frac{M^2-M+2}{M}\sum_{i \in V_c\cup V_d}\ket{i}-2\sum_{i \in V_e}\ket{i}-\sum_{i \in V_f}\ket{i}\right]\xrightarrow[]{\text{O.N.}} \ket{e_4}\,,\\
H\ket{e_4}=&\frac{1}{\sqrt{(M-1)(M^2-2M+4)(M^3+2M^2-8M+16)}}\left\lbrace\vphantom{\sum_{i\in V_f}} (M^2-4)\left[\sum_{i\in V_a} \ket{i}-(M-1)\ket{b}\right]\right.\nonumber\\
&\left.+2(M^2-M+2)\sum_{i\in V_c\cup V_d} \ket{i} -M(M^2-2M+8)\sum_{i\in V_e} \ket{i}+(M-2)^2\sum_{i\in V_f} \ket{i}\right\rbrace\xrightarrow[]{\text{O.N.}} \ket{e_5}\,,
\end{align}
and it can be proved that

\begin{equation}
H\ket{e_5}=\frac{M+2}{M^3+2M^2-8M+16}\left[M\sqrt{(M-2)(M^2-2M+4)}\ket{e_4}+(M^3-4M+8)\ket{e_5}\right]\,.
\end{equation}

In conclusion, any state $H^k \vert w \rangle \in \operatorname{span}(\left\lbrace \vert e_1 \rangle, \ldots, \vert e_5\rangle \right\rbrace) \forall k \in \mathbb{N}_0$, thus the states \eqref{eq:basis_simplexCG} form an orthonormal basis for the subspace $\mathcal{I}(H,\ket{w})$.

%%%%%%%%%%%%%%%%%%%%%%%%%%%%%%%%%%%%%%%%%%
\reftitle{References}
\externalbibliography{yes}
\bibliography{biblio_tectqwg}

% The following MDPI journals use author-date citation: Arts, Econometrics, Economies, Genealogy, Humanities, IJFS, JRFM, Laws, Religions, Risks, Social Sciences. For those journals, please follow the formatting guidelines on http://www.mdpi.com/authors/references
% To cite two works by the same author: \citeauthor{ref-journal-1a} (\citeyear{ref-journal-1a}, \citeyear{ref-journal-1b}). This produces: Whittaker (1967, 1975)
% To cite two works by the same author with specific pages: \citeauthor{ref-journal-3a} (\citeyear{ref-journal-3a}, p. 328; \citeyear{ref-journal-3b}, p.475). This produces: Wong (1999, p. 328; 2000, p. 475)

%%%%%%%%%%%%%%%%%%%%%%%%%%%%%%%%%%%%%%%%%%
%% optional
%\sampleavailability{Samples of the compounds ...... are available from the authors.}

%% for journal Sci
%\reviewreports{\\
%Reviewer 1 comments and authors’ response\\
%Reviewer 2 comments and authors’ response\\
%Reviewer 3 comments and authors’ response
%}

%%%%%%%%%%%%%%%%%%%%%%%%%%%%%%%%%%%%%%%%%%
\end{document}